\documentclass{aa}  

\usepackage{graphicx}
\usepackage{natbib}
\usepackage{scalerel}
\usepackage{hyperref}
\bibpunct{(}{)}{;}{a}{}{,} 
\usepackage[varg]{txfonts}
\usepackage{multirow}
\usepackage{booktabs,tabularx}
\begin{document}

\newcommand{\vdag}{(v)^\dagger}
\newcommand{\fdw}{$\rm f_{dg}$}
\newcommand{\tioii}{$\rm TiO2_{SDSS}$}
\newcommand{\nad}{$\rm NaD$}
\newcommand{\dtioii}{$\rm \delta(TiO2_{SDSS})$}
\newcommand{\dnad}{$\rm \delta(NaD)$}
\newcommand{\lcdm}{$\Lambda$CDM }
\newcommand{\morgana}{{\sc morgana }}
\newcommand{\munich}{{\it Munich }}\newcommand{\somer}{SC-SAM }
\newcommand{\msun}{{\rm M}_\odot}
\newcommand{\msunyr}{{\rm M}_\odot\ {\rm yr}^{-1}}
\newcommand{\gaea}{\sc{gaea}}
\def\lesssim{\lower.5ex\hbox{$\; \buildrel < \over \sim \;$}}
\def\gtrsim{\lower.5ex\hbox{$\; \buildrel > \over \sim \;$}}

\title{Variation of the stellar initial mass function in
  semi-analytical models III: testing the cosmic ray regulated
  integrated galaxy-wide initial mass function}

\author{Fabio Fontanot\inst{1,2}
\and Francesco La Barbera\inst{3}
\and Gabriella De Lucia\inst{1,2}
\and Rachele Cecchi\inst{4}
\and Lizhi Xie\inst{5}
\and Michaela Hirschmann\inst{6,1}
\and Gustavo Bruzual\inst{7}
\and St\'ephane Charlot\inst{8}
\and Alexandre Vazdekis\inst{9,10}
}

\institute{INAF - Astronomical Observatory of Trieste, via G.B. Tiepolo 11, I-34143 Trieste, Italy (fabio.fontanot@inaf.it)\label{inst1}
\and IFPU - Institute for Fundamental Physics of the Universe, via Beirut 2, 34151, Trieste, Italy\label{inst2}
\and INAF - Astronomical Observatory  of Capodimonte, sal. Moiariello, 16, I-80131, Napoli, Italy\label{inst3}
\and Astronomy Sector, Department of Physics, University of Trieste, via Tiepolo 11, 34143, Trieste, Italy \label{inst4}
\and Tianjin Astrophysics Center, Tianjin Normal University, Binshuixidao 393, 300384, Tianjin, China\label{inst5}
\and Institute for Physics, Laboratory for Galaxy Evolution, EPFL, Observatoire de Sauverny, Chemin Pegasi 51, 1290 Versoix, Switzerland \label{inst6}
\and Instituto de Radioastronom\'\i a y Astrof\'\i sica, UNAM, Campus Morelia, C.P. 58089, Morelia, M\'exico\label{inst7}
\and Sorbonne Universit\'e, CNRS, UMR 7095, Institut d'Astrophysique de Paris, 98 bis bd Arago, 75014 Paris, France\label{inst8}
\and Instituto de Astrof\'isica de Canarias, E-38200 La Laguna, Tenerife, Spain\label{inst9}
\and Departamento de Astrof\'isica, Universidad de La Laguna, E-38205 La Laguna, Tenerife, Spain\label{inst10}
}

\date{Accepted ... Received ...}

\abstract  
  {In our previous work, we derive the CR-IGIMF: a new scenario for a variable stellar initial mass function (IMF), which combines numerical results on the role played by cosmic rays in setting the thermal state of star forming gas, with the analytical approach of the integrated galaxy-wide IMF.}
  {In this work, we study the implications of this scenario for the properties of local Early-Type galaxies (ETG), as inferred from dynamical, photometric and spectroscopic studies.}
  {We implement a library of CR-IGIMF shapes in the framework of the Galaxy Evolution and Assembly ({\gaea}) model. {\gaea} provides predictions for the physical and photometric properties of model galaxies, and for their chemical composition. Our realization includes a self-consistent derivation of synthetic spectral energy distribution for each model galaxy, allowing a direct derivation of the mass fraction in the mean IMF of low-mass stars (i.e. the dwarf-to-giant ratio - $\rm f_{dg}$) and a comparison with IMF sensitive spectral features.}
  {The predictions of the {\gaea} model implementing the CR-IGIMF confirm our previous findings: it correctly reproduces both the observed excess of z$\sim$0 dynamical mass (mass-to-light ratios) with respect to spectroscopic (photometric) estimates assuming a universal, MW-like, IMF, and the observed increase of [$\alpha$/Fe] ratios with stellar mass in spheroidal galaxies. Moreover, this realization reproduces the increasing trends of $\rm f_{dg}$, and IMF-sensitive line-strengths with velocity dispersion, although the predicted relations are significantly shallower than the observed ones.}
  {Our results show that the CR-IGIMF is a promising scenario that reproduces {\it at the same time} dynamical, photometric and spectroscopic indications of a varying IMF in local ETGs. The shallow relations found for spectral indices suggest that either a stronger variability as a function of galaxy properties or additional dependences (e.g. as a function of star forming gas metallicity) might be required to match the strength of the observed trends.}


\titlerunning{\footnotesize CR-IGIMF in {\gaea}}
\authorrunning{Fontanot et al.}
   \maketitle
%

\section{Introduction}\label{sec:intro}                                       
The stellar initial mass function (IMF) represents a fundamental
ingredient for theoretical models of galaxy formation and
evolution. For each given star formation episode, this statistical
function defines the number of stars formed per stellar mass bin and
therefore it determines the number of SNe and the fraction of baryonic
mass locked in long-living stars. Direct observations of star-forming
regions in the Milky Way (MW) show a remarkable consistency in the
derived shape of the IMF. Different authors proposed different
functional shapes for the IMF in our galaxy \citep{Salpeter55,
  Kroupa01, Chabrier03}, that mainly differ in the very uncertainty
regime of brown dwarfs, i.e. at the low-mass end. Given the fact that
the MW is the only galaxy where we can currently measure the IMF using
direct stellar countings, the near invariance of the results prompted
the idea of its universality. However, from a theoretical
perspective, the physical description of gravitational collapse and
fragmentation of molecular clouds (MCs) is still debated in the
literature \citep[see e.g.][]{Krumholz14}: several models \citep[among
  others]{Klessen05, WeidnerKroupa05, HennebelleChabrier08,
  Papadopoulos11, NarayananDave13} explored the range of possible IMF
shapes depending on the physical properties of the interstellar medium
(ISM).

A first indication that the IMF can indeed vary in environments
different from those of the MW disc, come from \citet{Klessen07}, who
show that the densest regions of the MW bulge (where stellar counts
are still accessible) are better described by a different IMF
(i.e. with a larger fraction of massive stars) with respect to the
star forming regions in the disc. Nonetheless, the assumption of a
{\it universal} IMF\footnote{In the following we will consider as a
reference universal IMF a Kroupa/Chabrier IMF, i.e. a multi-slope
broken power-law, with a steeper high-mass end and a break at around 1
M$_\odot$.} is still used in almost all extragalactic studies. The
main limitation for the study of the IMF shape over a variety of
environments is our inability to resolve individual star clusters in
external galaxies. Therefore, the only information currently
accessible comes from the integrated light from composite stellar
populations, binding our ability to detect signatures of deviations
from a universal IMF to indirect evidences.

The number of evidences in favour of a varying IMF have steadily
increased in recent times. Using galaxies from the
GAMA\footnote{Galaxy And Mass Assembly} survey, \citet{Gunawardhana11}
show that their optical colours and H$_\alpha$ line strengths suggest a
flatter high-mass slope of the IMF at increasing SFR. Dynamical
studies in a sample of early-type galaxies (ETGs) in the ATLAS$^{\rm
  3D}$ survey have revealed a systematic excess of dynamical
mass-to-light ratios with respect to the expectations for a universal,
MW-like, IMF based on observed photometry. This mass excess, $\alpha$,
increases with increasing galaxy velocity dispersion $\sigma$
\citep{Cappellari12}. A number of spectral features in the spectra of
ETGs (such as Na$_{\rm I}$, TiO1 and TiO2), sensitive to temperature
and/or surface gravity, have been proposed as tracers of the ratio
between low-mass and giant stars, and therefore of the shape of the
IMF they are formed from. Several groups \citep[see][among the
  others]{ConroyvanDokkum12, Ferreras13, LaBarbera13, Spiniello14}
have adopted this approach to conclude that the IMF slope at the
low-mass end becomes steeper at increasing $\sigma$ and/or stellar
mass M$_\star$.

The main problem in interpreting this wealth of information in a
consistent picture lies in our lack of physical understanding on how
the IMF shape should vary as a function of the physical conditions of
star forming regions. IMFs with more low-mass or high-mass stars than
the MW-like IMF are defined ``bottom-heavy'' (or ``top-light'') and
``top-heavy'' (or ``bottom-light''), respectively. Spectroscopic
studies seem to prefer a ``bottom-heavy'' solution for more massive
galaxies (i.e. an IMF slope at the low-mass end steeper than the
MW-like IMF). However, these studies are based only on the light of
evolved stars, and therefore this approach cannot provide any
constraint on the high-mass end of the IMF. In principle, the
dynamical technique is sensitive to the overall shape of the IMF,
however, it cannot distinguish between a ``bottom-heavy'' or a
``top-heavy'' scenario \citep{Tortora13}. In fact, the observed
$\alpha$ excess of the mass-to-light ratio can be interpreted either
as a larger fraction of low-mass stars (giving rise to direct mass
excess) or as a larger fraction of massive short-lived stars (giving
rise to a decrease in the total light at late times).

The situation is further complicated by the detection of radial IMF
gradients in ETGs \citep{MartinNavarro15, LaBarbera17, vanDokkum17,
  Sarzi18, LaBarbera21}. The results suggest that the low-mass end of the IMF
becomes steeper towards the inner galaxy regions, while being
comparable with a MW-like IMF in the outer regions. However, the same
galaxies are more $\alpha$ enriched in the inner regions than in the
outskirts, and these [$\alpha$/Fe] gradients may be better explained
by a larger number of Type II SNe in the center (thus suggesting an
IMF becoming more ``top-heavy'', towards the central regions). A
possible solutions to these tensions in the interpretation of the
available data may be a time-varying IMF, where the two ends are
allowed to vary independently over different time-scales
\citep[see e.g.][]{Weidner13, Ferreras15}. The fact that the two
slopes can evolve in different directions at the same time makes it
difficult to recast the differences with respect to the MW-like IMF
into the standard definition of ``bottom-heavy'' or ``top-heavy''
IMF. In the following, we use this nomenclature only when discussing
predictions related to one of the ends.

Galaxy formation models that self-consistently follow the interplay
between IMF and galaxy properties represents a key tool to assess the
complexities in the interpretation of the photometric/spectroscopic
data. In our previous work, on the variable IMF hypothesis, we tested
two different models, namely the integrated galaxy-wide IMF (IGIMF) by
\citet{WeidnerKroupa05} and the Cosmic Ray (CR) regulated IMF from
\citet[][PP11 hereafter]{Papadopoulos11}. In brief, the former
approach estimates the galaxy-wide IMF starting from a limited number
of physically and observationally motivated axioms that can be recast
as a function of the galaxy SFR. Instead, PP11 provide numerical
estimates for the characteristic Jeans mass of young stars
($M^{\star}_{\rm J}$) forming in a molecular cloud (MC) embedded in a
CR energy density field of a given intensity. As energetic CRs can
penetrate deeper into the core of these MCs, they can change the
chemical and thermal state of the star forming gas, thus affecting the
emerging IMF. In this case, the dependence can be recast as a function
of the SFR surface density\footnote{In \citet{Fontanot18a} we assume
that the SFR surface density is a good tracer of the CR energy
density.}. In \citet[F17]{Fontanot17a} and \citet[F18a]{Fontanot18a}
we present theoretical models implementing the IGIMF and PP11
approaches, respectively, and discuss their implications for galaxy
formation and evolution. These models are able to reproduce both the
[$\alpha$/Fe] versus stellar mass relation and the $\alpha$ excess of
model ETGs at z$\sim$0. Nonetheless, the typical IMF variations in
both these models mainly affect only the slope of the high-mass end,
while the low-mass end of the IMF is constant by
construction. Therefore, these models are not able to reproduce the
variation of the low-mass end slope derived from the spectroscopic
studies of nearby ETGs.

In order to overcome this limitation, \citet[F18b
  hereafter]{Fontanot18b} proposed a new IMF derivation (dubbed
CR-IGIMF), based on a combination of the two previous approaches. The
main feature of the CR-IGIMF is the predicted evolution of both the
high-mass and low-mass ends of the IMF, as a function of the physical
properties of model galaxies. In F18b, we presented a preliminary
analysis of the impact of this scenario on the spectral properties of
galaxies, based on idealized star formation histories, that we used to
estimate the mean IMF shapes of galaxies at z$\sim$0 as a function of
their stellar mass. We concluded that the CR-IGIMF qualitatively
reproduces the observed trends. The aim of this work is to explore in
a more quantitative way the impact of this new IMF model on galaxy
evolution, and in particular to explore the constraints on dynamical
and spectroscopic properties of ETGs at $z\sim0$. To this end, we
implement the CR-IGIMF scenario in the framework of the semi-analytic
model for GAlaxy Evolution and Assembly ({\gaea}), and employ the
synthetic stellar population models of \citet{Vazdekis12}, to compute
synthetic spectra of model galaxies. These implementations provide a
new, self-consistently calibrated, tool to fully describe the IMF
evolution along the star formation and assembly history of model
galaxies.

This paper is organized as follows. In Section~\ref{sec:compimf} we
layout the theoretical framework of the CR-IGIMF derivation, as
presented in F18b. In Section~\ref{sec:gaea}, we introduce the
implementation of the CR-IGIMF in the {\gaea} model. We present our
main results and discuss them in the context of galaxy formation and
evolution in Section~\ref{sec:results}. Finally, we outline our main
conclusions in Section~\ref{sec:final}.

\section{CR-IGIMF}\label{sec:compimf}
\begin{table*}
  \caption{Analytical fits to the CR-IGIMF.}
  \label{tab:fits}
  \renewcommand{\footnoterule}{}
  \centering
  \begin{tabular}{cc|ccccccccc|ccc}
    \hline
    SFR & $\Sigma_{\rm SFR} / \Sigma_{\rm MW}$ & $\alpha_{\rm I}$ & m$_{\rm I}$ & $\alpha_{\rm II}$ & m$_{\rm II}$ & $\alpha_{\rm III}$ & m$_{\rm III}$ & $\alpha_{\rm IV}$ & m$_{\rm IV}$ & $\alpha_{\rm V}$ & $\rm f^{\star}_{dg}$ & $(1-\rm f_{ret})^{\star}$ & $\rm f^{\star}_{SN2}$  \\
    (M$_\odot$ yr$^{-1}$) & & & (M$_\odot$) & & (M$_\odot$) & & (M$_\odot$) & & (M$_\odot$) & & & \\
    \hline
    10$^-3$ &   0.1  & -1.28 &  0.31  &  -1.35  &   1.30 &  -2.92 &   5.18 &  -3.70 &  20.99 &  -14.65  &  0.89 & 0.95 & 0.27  \\
     0.01   &        & -1.42 &  0.34  &  -1.63  &   1.38 &  -2.81 &   7.96 &  -3.46 &  55.94 &  -14.64  &  0.96 & 1.00 & 0.39  \\
      0.1   &        & -1.34 &  0.21  &  -1.72  &   1.25 &  -2.46 &   1.89 &  -2.79 &  76.84 &   -4.10  &  1.01 & 1.03 & 0.58  \\
        1   &        & -1.54 &  0.21  &  -1.80  &   1.48 &  -2.73 &   5.11 &  -2.57 &  20.04 &   -2.40  &  1.03 & 1.02 & 0.70  \\
       10   &        & -1.84 &  0.25  &  -1.97  &   1.50 &  -2.42 &   6.52 &  -2.20 &  25.23 &   -2.05  &  1.08 & 0.96 & 1.06  \\
      100   &        & -1.98 &  0.24  &  -2.21  &   0.45 &  -2.09 &   6.88 &  -1.90 &  24.31 &   -1.79  &  1.13 & 0.82 & 1.51  \\
    10$^3$  &        & -2.16 &  0.25  &  -2.27  &   0.97 &  -1.83 &   5.18 &  -1.67 &  21.74 &   -1.57  &  1.17 & 0.61 & 1.97  \\
\hline                                                                                                                                                                      
    10$^-3$ &    1   & -1.30 &  0.45  &  -1.37  &   1.52 &  -3.10 &   9.33 &  -4.19 &  21.98 &  -17.42  &  0.89 & 0.91 & 0.29  \\
     0.01   &        & -1.30 &  0.25  &  -1.54  &   1.36 &  -2.85 &  11.46 &  -3.59 &  56.34 &  -14.73  &  0.95 & 0.96 & 0.42  \\
      0.1   &        & -1.30 &  0.27  &  -1.71  &   1.32 &  -2.50 &   2.13 &  -2.80 &  76.99 &   -4.13  &  0.98 & 1.00 & 0.61  \\
        1   &        & -1.51 &  0.26  &  -1.79  &   1.58 &  -2.74 &   5.09 &  -2.59 &  20.38 &   -2.41  &  1.02 & 0.99 & 0.73  \\
       10   &        & -1.80 &  0.31  &  -1.95  &   1.58 &  -2.42 &   6.80 &  -2.22 &  24.39 &   -2.06  &  1.08 & 0.94 & 1.08  \\
      100   &        & -2.00 &  0.27  &  -2.09  &   6.33 &  -1.96 &  13.00 &  -1.87 &  31.57 &   -1.79  &  1.12 & 0.81 & 1.53  \\
    10$^3$  &        & -2.16 &  0.34  &  -2.27  &   0.97 &  -1.83 &   5.20 &  -1.67 &  21.64 &   -1.57  &  1.17 & 0.61 & 1.98  \\
\hline                                                                                                                                                   
    10$^-3$ &   10   & -1.30 &  1.16  &  -2.02  &   3.34 &  -3.29 &  13.08 &  -4.95 &  22.67 &  -19.82  &  0.88 & 0.76 & 0.52  \\ 
     0.01   &        & -1.30 &  0.57  &  -1.50  &   1.25 &  -2.22 &   5.20 &  -3.50 &  55.67 &  -14.37  &  0.90 & 0.81 & 0.63  \\
      0.1   &        & -1.30 &  0.61  &  -1.68  &   1.23 &  -2.09 &   3.05 &  -2.85 &  77.40 &   -4.12  &  0.91 & 0.81 & 0.81  \\
        1   &        & -1.50 &  0.71  &  -1.73  &   1.25 &  -2.14 &   3.13 &  -2.69 &  23.52 &   -2.45  &  0.95 & 0.83 & 0.92  \\
       10   &        & -1.75 &  0.97  &  -2.01  &   3.07 &  -2.46 &   8.00 &  -2.31 &  24.39 &   -2.10  &  1.02 & 0.81 & 1.24  \\
      100   &        & -1.99 &  0.70  &  -1.95  &   3.05 &  -2.12 &   8.80 &  -1.95 &  26.45 &   -1.81  &  1.09 & 0.72 & 1.63  \\
    10$^3$  &        & -2.08 &  0.63  &  -2.20  &   1.04 &  -1.79 &   9.00 &  -1.66 &  25.90 &   -1.57  &  1.14 & 0.57 & 2.03  \\
\hline                                                                                                                                                   
    10$^-3$ &  100   & -1.30 &  0.26  &  -1.28  &   1.27 &  -2.26 &   9.42 &  -4.53 &  22.30 &  -18.68  &  0.87 & 0.71 & 0.95  \\
     0.01   &        & -1.30 &  0.43  &  -1.34  &   1.66 &  -2.38 &  11.39 &  -3.87 &  58.87 &  -18.11  &  0.89 & 0.67 & 1.11  \\
      0.1   &        & -1.30 &  0.33  &  -1.32  &   1.51 &  -2.22 &   7.08 &  -2.94 &  78.04 &   -4.24  &  0.88 & 0.62 & 1.25  \\
        1   &        & -1.30 &  0.35  &  -1.47  &   1.64 &  -2.20 &   7.05 &  -2.78 &  25.12 &   -2.53  &  0.92 & 0.64 & 1.35  \\
       10   &        & -1.30 &  0.35  &  -1.67  &   2.06 &  -2.12 &   7.25 &  -2.43 &  24.79 &   -2.15  &  0.96 & 0.62 & 1.62  \\
      100   &        & -1.30 &  0.35  &  -1.96  &   0.90 &  -1.60 &   2.10 &  -2.03 &  24.89 &   -1.85  &  1.02 & 0.56 & 1.93  \\
    10$^3$  &        & -1.30 &  0.35  &  -2.15  &   0.82 &  -1.74 &  18.08 &  -1.64 &  36.40 &   -1.58  &  1.05 & 0.45 & 2.23  \\
\hline                                                                                                                                                   
    10$^-3$ & 10$^3$ & -1.30 &  1.37  &  -2.31  &   9.37 &  -3.20 &  18.75 &  -8.04 &  24.44 &  -37.78  &  0.88 & 0.71 & 1.12  \\ 
     0.01   &        & -1.30 &  1.35  &  -2.17  &  13.37 &  -3.83 &  50.00 &  -7.68 &  65.52 &  -40.60  &  0.88 & 0.62 & 1.46  \\
      0.1   &        & -1.30 &  1.18  &  -1.89  &   5.47 &  -2.42 &  13.20 &  -3.00 &  77.98 &   -4.40  &  0.88 & 0.56 & 1.70  \\
        1   &        & -1.30 &  1.20  &  -1.89  &   5.79 &  -2.41 &  15.85 &  -2.94 &  25.35 &   -2.59  &  0.88 & 0.54 & 1.79  \\
       10   &        & -1.30 &  1.25  &  -1.78  &   5.73 &  -2.24 &  16.22 &  -2.51 &  26.27 &   -2.20  &  0.88 & 0.45 & 2.09  \\
      100   &        & -1.30 &  1.30  &  -1.70  &   5.22 &  -2.00 &  13.44 &  -2.04 &  30.19 &   -1.87  &  0.88 & 0.37 & 2.34  \\
    10$^3$  &        & -1.30 &  1.32  &  -1.62  &   4.82 &  -1.74 &  23.13 &  -1.64 &  43.16 &   -1.59  &  0.88 & 0.29 & 2.52  \\
\hline                                                                                                                                                   
    10$^-3$ & 10$^4$ & -1.30 &  1.41  &  -2.34  &  10.67 &  -3.45 &  19.35 &  -7.39 &  24.35 &  -34.27  &  0.88 & 0.71 & 1.11  \\
     0.01   &        & -1.30 &  1.31  &  -2.13  &  12.65 &  -2.40 &  31.59 &  -4.91 &  62.55 &  -26.49  &  0.88 & 0.59 & 1.79  \\
      0.1   &        & -1.30 &  1.18  &  -2.02  &   4.00 &  -2.00 &  27.97 &  -3.12 &  73.28 &   -4.39  &  0.88 & 0.52 & 2.08  \\
        1   &        & -1.30 &  1.19  &  -1.94  &  11.94 &  -1.97 &  27.27 &  -2.77 &  77.57 &   -3.19  &  0.88 & 0.48 & 2.21  \\
       10   &        & -1.30 &  1.19  &  -1.74  &   2.09 &  -1.68 &   6.29 &  -1.85 &  26.43 &   -2.39  &  0.88 & 0.37 & 2.56  \\
      100   &        & -1.30 &  1.14  &  -1.48  &   6.18 &  -1.75 &  26.88 &  -2.06 &  44.84 &   -1.98  &  0.88 & 0.30 & 2.75  \\ 
    10$^3$  &        & -1.30 &  1.18  &  -1.36  &   5.91 &  -1.62 &  27.85 &  -1.72 &  44.22 &   -1.65  &  0.88 & 0.24 & 2.77  \\
\hline     
           
  \end{tabular}
  \vspace{1ex}

  The quantities $\rm f^{\star}_{dg}$, (1-$\rm f_{ret}$)$^{\star}$ and
  $\rm f^{\star}_{SN2}$ represent, respectively, the value of the
  dwarf-to-giant ratio, of the mass fraction locked in low-mass stars
  and of the Type II SNe fraction, relative to the same quantities
  computed for a MW-like IMF with a \citet{Kroupa01} multi-slope
  functional form.
\end{table*}
In this paper, we take advantage of the recent derivation of the IGIMF
$\varphi_{\rm IGIMF}$ as proposed in \citet{Fontanot18b}, which
accounts for the effect of the Cosmic Rays on the thermal and chemical
properties of star-forming molecular clouds. This approach is similar
to the one proposed by \citet[see also
  \citealt{Dib23}]{WeidnerKroupa05} and it is based on the {\it
  ansatz} of integrating the individual IMFs associated with each MC
($\varphi_\star(m)$), weighted by the MC mass function $\varphi_{\rm
  CL}(M_{\rm cl})$:

\begin{equation}
\varphi_{\rm IGIMF}(m) = \int^{M_{\rm cl}^{\rm max}}_{M_{\rm cl}^{\rm
    min}} \varphi_\star(m \le m_\star^{\rm max} (M_{\rm cl}))
\varphi_{\rm CL}(M_{\rm cl}) dM_{\rm cl}
\label{eq:igimf}
\end{equation}

\noindent
Eq.~\ref{eq:igimf} critically depends both on the maximum value of the
mass of a star cluster to form ($M_{\rm cl}^{\rm max}$) and on the
largest stellar mass forming in a given cluster ($m_{\rm max}$). Using
available evidence from the Taurus-Auriga complex
\citep{KroupaBouvier03}, we set $M_{\rm cl}^{\rm min}=5 M_\odot$ as
the minimum mass for star clusters. Additional {\it ansatzs} are then
used to define the key quantities entering Eq.~\ref{eq:igimf} as a
function of the assumed SFR.

\begin{equation}
\log \frac{{\rm M}_{\rm cl}^{\rm max}}{\msun} = 0.746 \, \log \frac{\rm SFR}{\msunyr}  +4.93;
\label{eq:mclmax}
\end{equation}

\noindent
Eq.~\ref{eq:mclmax} describes the scaling of $M_{\rm cl}^{\rm max}$
with SFR, and it has been derived by \citet{Weidner04} using stellar
data. Consistently with F17 and F18a, we assume $M_{\rm cl}^{\rm max}
\leq 2 \times 10^7 \msun$.

\begin{equation}
\log \frac{m_\star^{\rm max}}{\msun} = 2.56 \log \frac{{\rm M}_{\rm cl}}{\msun} \left[ 3.82^{9.17} + (\log \frac{{\rm M}_{\rm cl}}{\msun})^{9.17} \right]^{1/9.17}-0.38
\label{eq:msmax}
\end{equation}

\noindent
Eq.~\ref{eq:msmax} is used to predict the maximum stellar mass
($m_\star^{\rm max}$) to form in a cluster of mass $M_{\rm
  cl}$. \citet{PflammAltenburg07} derive it assuming that each stellar
cluster contains exactly one $m_{\rm max}$ star.

\begin{equation}
\varphi_{\rm CL}(M_{\rm cl}) \propto M_{\rm cl}^{-\beta};
\label{eq:clmf}
\end{equation}

\noindent
Eq.~\ref{eq:clmf} represents the functional shape for the star cluster
mass function.

\begin{equation}
\beta = \left\{
\begin{array}{ll}
2 & {\rm SFR} < 1 \msunyr \\
-1.06 \, \log \frac{\rm SFR}{\msunyr} +2 & {\rm SFR} \ge 1 \msunyr \\
\end{array}
\right.
\label{eq:beta}
\end{equation}

\noindent
Eq.~\ref{eq:beta} provides the allowed values for the $\beta$
parameter in Eq.~\ref{eq:clmf}. \citet{LadaLada03} suggest $\beta=2$
based on observational results in the local Universe. The flattening
of $\beta$ at high-SFRs is suggested by \citet{Gunawardhana11} in
order to reproduce the stellar population of galaxies in the GAMA
survey.

\begin{equation}
\alpha_3 = \left\{
\begin{array}{ll}
2.35 & \rho_{\rm cl} < 9.5 \times 10^4 \, \msun/pc^3\\
1.86-0.43 \, \log(\frac{\rho_{\rm cl}}{10^4 \, \msun/pc^3}) & \rho_{\rm cl} \ge 9.5 \times 10^4 \, \msun/pc^3 \\
\end{array}
\right.
\label{eq:alpha3}
\end{equation}

\noindent
Eq.~\ref{eq:alpha3} quantifies the dependence of the high-mass end
($\alpha_3$) on the cluster core density ($\rho_{\rm cl}$) as proposed
in \citet{Marks12}. This deviation from a Salpeter high-mass end slope
has been reported in the literature (see \citealt{Kroupa13} for a
review of the subject). It is important to keep in mind that in
Eq.~\ref{eq:alpha3} we choose to focus on the dependence of $\alpha_3$
on $\rho_{\rm cl}$ and neglect any possible dependence of $\alpha_3$
on metallicity. Our choice is functional to provide a framework for
the IGIMF derivation based on SFR and SFR density. \citet{Jerabkova18}
explore an alternative derivation of IGIMF considering metallicity and
SFR as the main drivers of its variation, finding similar conclusions
as in F18b (see discussion below). Finally, the system of
  equations is closed by considering the link between the cluster core
  density and cluster mass as given by Eq.~\ref{eq:rhocl}
  \citep{MarksKroupa12}: 

\begin{equation}
  \log \frac{\rho_{\rm cl}}{\msun/pc^3} = 0.61 \, \log \frac{{\rm M}_{\rm cl}}{\msun} +2.85.
\label{eq:rhocl}
\end{equation}

In the original \citet{WeidnerKroupa05} framework, the IGIMF
associated with a given SFR event is numerically derived by assuming a
universal $\varphi_\star(m)$ in each individual star-forming
MC. Following our previous works, we assume that the IMF shape in each
MC is well described by the \citet{Kroupa01} multi-slope functional
form:

\begin{equation}
\varphi_\star(m) =  \left\{
\begin{array}{ll}
(\frac{m}{m_{\rm low}})^{-\alpha_1} & m_{\rm low} \le m < m_{\rm br} \\
(\frac{m_{\rm br}}{m_{\rm low}})^{-\alpha_1} (\frac{m}{m_{\rm br}})^{-\alpha_2} & m_{\rm br} \le m < m_1 \\
(\frac{m_{\rm br}}{m_{\rm low}})^{-\alpha_1} (\frac{m_1}{m_{\rm br}})^{-\alpha_2} (\frac{m}{m_1})^{-\alpha_3} & m_1 \le m \le m_{\rm max} \\
\end{array}
\right.
\label{eq:kroimf}
\end{equation}

\noindent
In Eq.~\ref{eq:kroimf} we already introduce the 3-slope formalism used
in F18b, by considering a variable inner break at $m_{\rm br}$; we
then fixed the remaining parameters at $m_{\rm low}=0.1 \, \msun$, $m_1=1.0 \, \msun$,
$\alpha_1=1.3$ and $\alpha_2=2.35$. The variable break $m_{\rm br}$
represents the novelty introduced by F18b, as it relaxes the
\citet{WeidnerKroupa05} {\it ansatz} of a universal IMF in individual
MCs\footnote{It is worth noting that variations in the IMF shape of
individual MCs have already been taken in account in the IGIMF
framework through Eq.~\ref{eq:alpha3} \citep[see also][]{Yan17}, based
on the local density of the MC and on the total SFR. However, the
results discussed in \citet{Papadopoulos11} show that IMF variations
can affect also its low-mass end shape. The F18b approach thus
accounts for both effects, at the same time.}. Instead, we assume that
stars in a given MCs form from an IMF well described by
Eq.~\ref{eq:kroimf}, where the position of the break at $m_{\rm br}$
depends on the characteristic Jeans mass of young stars
($M^{\star}_{\rm J}$), that in turn depends on the MC density and on
the environment where it lives, and in particular on the CR density
field. In detail, for a given CR energy density ($U_{\rm CR}$) and core
density given by Eq.~\ref{eq:rhocl}, $M^{\star}_{\rm J}$ is derived
using the numerical solutions to the chemical and thermal equations
regulating the CR-dominated ISM (i.e. Fig.~4 in P11):

\begin{equation}
  m_{\rm br} = M^{\star}_{\rm J}(\rho_{\rm cl},U_{\rm CR}).
  \label{eq:mbr}
\end{equation}

The numerical solutions we consider are characterized by an increase
of $M^{\star}_{\rm J}$ with increasing $U_{\rm CR}$, due to the higher
CR heating rate associated with the increased energy density. At fixed
$U_{\rm CR}$, PP11 show also that $M^{\star}_{\rm J}$ decreases at
increasing $\rho_{\rm cl}$. This trend may appear counter intuitive,
as high-mass MCs are expected to be hotter and the Jeans mass
increases with temperature. However, the numerical results suggest
that, when the contribution of CR energy density is taken into
account, the dependence of the Jeans mass on density is dominant with
respect to temperature.

\subsection{Library}

 Consistently with our previous work, in the following we employ a
  CR-IGIMF binned library (see Tab.~\ref{tab:fits}). The use of a
  predefined, limited, set of IMFs allows us to retain the
  computational efficiency of the semi-analytic approach without
  losing precision. We consider 7 equally spaced bins in $\log$(SFR)
  in the range [-3,3] and 6 $U_{\rm CR}$ values relative to the
  corresponding MW CR density ($U_{\rm MW}$, estimated from the MW SFR
  density). While the 6 $U_{\rm CR}/U_{\rm MW}$ are limited by the
  original computations in P11, we employ 7 bins in $\log$(SFR) after
  having tested the numerical convergence of broad band magnitudes
  predicted by the model (see Sec.~\ref{sec:runs}). We numerically
  integrate Eq.~\ref{eq:igimf} in each of the 42 total bins
  combinations, and we fit these numerical results with a
  multi-component power law, consistent with an extension of the
  Kroupa IMF formalism. We start from Eq.~\ref{eq:kroimf} and we then
  add slopes and break points until we get the required convergence of
  the fit (i.e. using an additional slope and break point results in a
  non-significant increase of the reduced $\chi^2$ probability). Our
  procedure indicates that a multi-component power law, with 5 slopes
  ($\alpha_{\rm I}$, $\alpha_{\rm II}$, $\alpha_{\rm III}$,
  $\alpha_{\rm IV}$, $\alpha_{\rm V}$, going from the faint end to the
  bright end) and 4 break points (m$_{\rm I}$, m$_{\rm II}$, m$_{\rm
    III}$, m$_{\rm IV}$, going from low mass to high mass) is required
  to fully describe the IMF shapes predicted by the CR-IGIMF. We
  collect the best fit values, describing the multi-component
  power-laws, in Table~\ref{tab:fits}: some of the most relevant
  CR-IGIMF shapes have been shown in Fig.~1 of F18b.

The possibility of an intrinsic IMF variation of individual MCs has a
fundamental impact on the shape of the emerging galaxy-wide IMF
predicted by the IGIMF framework.
\begin{itemize}
\item{For the typical conditions of a MW-like star forming region
  ($U_{\rm CR} \sim U_{\rm MW}$ and SFR$\sim$1$\msunyr$), the CR-IGIMF
  predicts a MW-like shape for the IMF, in agreement with local
  observations.}
\item{At fixed SFR, the low-mass end of the CR-IGIMF steepens as
  $U_{\rm CR}$ decreases; however, it becomes indistinguishable from a
  Kroupa-like $\alpha_1$ slope for $U_{\rm CR}$ larger than ten times
  the typical MW value. Indeed, the characteristic Jeans mass of young
  stars decreases as $U_{\rm CR}$ decreases (Fig.~4 in PP11), thus
  leading to smaller $m_{\rm br}$ (Eq.~\ref{eq:mbr}).}
\item{At fixed $U_{\rm CR}$, the high-mass end of the CR-IGIMF follows
  the typical IGIMF expectation, i.e. it becomes shallower at
  increasing SFR. This is driven by Eq.~\ref{eq:alpha3}, through the
  density of the cluster (following Eq.~\ref{eq:rhocl}).}
\item{At $U_{\rm CR}/U_{\rm MW} < 10$, the low-mass end of the
  CR-IGIMF steepens as SFR increases. This is due to the combined
  effect of the flattening of $\beta$ at high-SFR (Eq.~\ref{eq:beta}),
  with the decreasing of the characteristic Jeans mass of young stars
  in denser environments (Fig.~4 in PP11).}
\item{If $U_{\rm CR}/U_{\rm MW}>100$, the low-mass end slope is again
  indistinguishable from an $\alpha_1$ slope. Indeed, in this $U_{\rm
    CR}$ range, $m_{\rm br}$ is always larger than 1 $\msun$ at all
  $\rho_{\rm cl}$ scales.}
\end{itemize}

The combination of the last two items implies that, at least for
$U_{\rm CR}$ smaller than 10 times the MW environment, a change in the
SFR affects both the high-mass and low-mass end slope at the same
time. In particular, an increase of the SFR above the typical MW-like
values flattens the high-mass end slope and steepens the low-mass end
slope of the CR-IGIMF at the same time, leading to a change in the
concavity of the IMF, for SFR$>100$ $\msunyr$.

Assuming a variable IMF affects different observable properties of
galaxies, as well as their evolution. In order to better quantify this
effect we report in Table~\ref{tab:fits} the values of some key simple
stellar population (SSP) properties\footnote{In order to estimate
these quantities, we compute integrals over the IMF shapes normalized
to unit solar mass formed stellar mass in appropriate mass ranges. In
particular, we compute \fdw as the ratio of the values obtained
integrating the IMF from 0.2 up to 0.5 $\msun$ and from 0.2 to 1
$\msun$ respectively; $\rm f_{SNII}$ represents the number of stars
larger than 8 $\msun$; we derive $\rm f_{ret}$ from the integral of
the total returned mass (along the entire stellar lifetime) as a
function of initial stellar mass.} such as the dwarf-to-giant ratio
($\rm f_{dg}$), the mass locked into long-lived stars (defined as the
complementary of the returned fraction $\rm f_{ret}$) and the Type II
SNe fraction ($\rm f_{SNII}$) per unit solar mass formed. In
Table~\ref{tab:fits}, we report the ratio of these quantities to the
corresponding values for a MW-like IMF with a \citet{Kroupa01}
functional form (i.e. Eq.~\ref{eq:kroimf}). The fraction of Type II
SNe, $\rm f_{SNII}$, shows the largest spread both in the entire
library and at fixed $U_{\rm CR}$, ranging from more than twice to
roughly one third of the expected value for a Kroupa IMF. This spread
is driven by the predicted evolution of the high mass end slope. The
mass locked in long-lived stars and stellar remnants decreases almost
monotonically at increasing SFR (due to the $\alpha_3$ evolution) and
$U_{\rm CR}$, reflecting mostly the increased number of high-mass
stars predicted by the CR-IGIMF. Finally, $\rm f_{dg}$ is
systematically lower than the Kroupa IMF expectation for $U_{\rm CR}
\gtrsim 100 U_{\rm MW}$ (due to $m_{\rm br}$ being always larger than
1 $\msun$ at all $\rho_{\rm cl}$ scales), while showing a wider range
of values for lower densities.  It is worth stressing that, by
construction, the maximum and minimum values for the low-mass end
slope cannot exceed the assumed $\alpha_1$ and $\alpha_2$,
  and the latter case implies a $\rm f_{dg}$ value 1.18 larger that
  corresponding to the Kroupa IMF (see also Sec.~\ref{sec:fdg}). These
values have been set by local observations of individual clouds and
are consistent with theoretical expectations \citep[see
  e.g][]{HennebelleChabrier08}. It is important to keep in mind that
it is unclear if the IMFs of individual MCs should respect these
constraints in all physical environments, especially in case strong CR
external fields are present (like the typical condition in starburst
galaxies, \citealt{Zhang18}).

As we already mentioned, \citet[see also \citealt{Yan21} for a
    more recent extension of the formalism]{Jerabkova18} propose an
alternative approach for an expansion of the IGIMF framework, which
considers a secondary dependence on metallicity (alongside with
SFR). Their results are qualitatively consistent with the scenario
considered in this work, the main differences lying at the low-mass
end. In particular, in the CR-IGIMF scenario the low-mass end slopes
steepen at decreasing SFR densities, and it equals the Kroupa IMF
value (i.e. $\alpha_1$) at $\Sigma_{\rm SFR} \gtrsim$ 100 $\Sigma_{\rm
  MW}$.  On the other hand, in \citet{Jerabkova18} the galaxy-wide IMF
(their model IGIMF3) can either be bottom-light or bottom-heavy if for
sub-Solar or super-Solar metallicities, respectively. Both approaches
associate a shallower high-mass end slope (corresponding to a larger
number high-mass stars with respect to the canonical $\alpha_2$ slope)
to $SFR>1\msunyr$ events. Finally, it is worth stressing that none of
these IGIMF derivations predicts, by construction, low-mass end slopes
steeper than $\alpha_2$.

\section{Semi-analytic model}\label{sec:gaea}
\begin{table}
  \caption{Parameter Calibration chart.}
  \label{tab:parameters}
  \renewcommand{\footnoterule}{}
  \centering
  \begin{tabular}{lc}
    \hline
    \multicolumn{2}{c}{Stellar Feedback parameters}  \\
    \hline
    $\alpha_{\rm SF}$            & 0.14   \\
    $\epsilon_{\rm reheat}$       & 0.4  \\
    $\epsilon_{\rm eject}$        & 0.15  \\
    $\kappa_{\rm radio}/10^{-5}$   & 1.4   \\
    $\gamma_{\rm reinc}$          & 1.0   \\
    \hline
  \end{tabular}
\end{table}

In this paper, we discuss the implications of the CR-IGIMF in a
cosmological framework, by implementing it in the state-of-the-art
{\gaea} semi-analytic model (SAM).  SAMs represent a modelling
technique aimed at describing the redshift evolution of galaxy
populations across cosmic times.  The physical processes responsible
for the energy and mass exchanges among the different baryonic
components are treated by means of a system of differential
equations. Specifically, each process is described using analytical
prescriptions that could be either empirically, numerically or
theoretically motivated. Several of these prescriptions require the
calibration of a fixed number of free parameters against a selected
set of observational constraints. Observational data outside the
calibration set then provide genuine predictions. When coupled with a
statistical realization of the evolution of Large Scale Structure as
traced by Dark Matter Haloes (i.e. a merger tree), SAMs represent a
flexible tool to study galaxy evolution in a cosmological volume: they
require only a fraction of the computational time usually associated
with hydro-simulations, while providing comparable representations of
the statistical properties of large galaxy samples. Moreover, the
limited computational request allows SAMs to sample their associated
multi-dimensional parameter space very efficiently. The main
disadvantage with respect to hydro-simulations lies in the intrinsic
difficulty to describe the distribution of baryons inside haloes, and
therefore to trace the spatial properties of galactic structures, like
discs and bulges. The coupling of the CR-IGIMF with a full fledged
theoretical model of galaxy evolution is a key improvement with
respect to our preliminary analysis in F18b, where we resort to
simplified assumptions and idealized star formation histories to have
an estimate of the IMF-sensitive spectral features. Indeed, the
availability of a self-consistently calibrated model allows us to
follow in detail the assembly of ETGs, by assigning a different IMF
shape from our library to each SSP that describes the galaxy
progenitors. This aspect is crucial in order to correctly establish
the contribution of each IMF to the final model ETG spectra.
\begin{figure}
  \includegraphics[width=9cm]{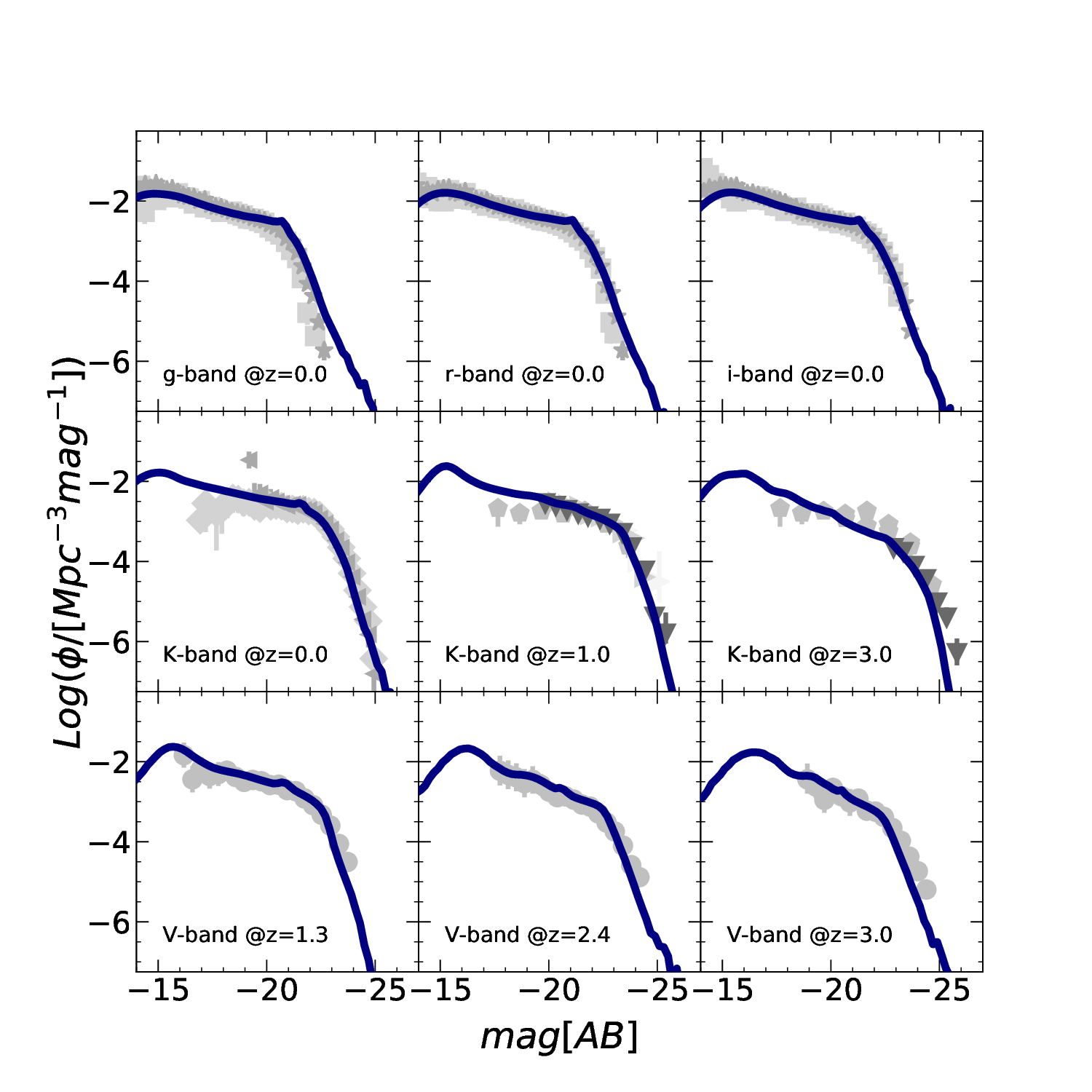} 
  \caption{Calibration set used for the CR-IGIMF run. Grey points
    represent the galaxy luminosity functions in different wavebands
    and at different redshifts including the SDSS g, r and i-band, K-
    and V-band (the same compilation of observational estimates used
    in F17 and F18a, see these papers for detailed references). In all
    panels, the blue line refers to the reference model
    predictions.}\label{fig:recal}
\end{figure}
\begin{figure}
  \includegraphics[width=9cm]{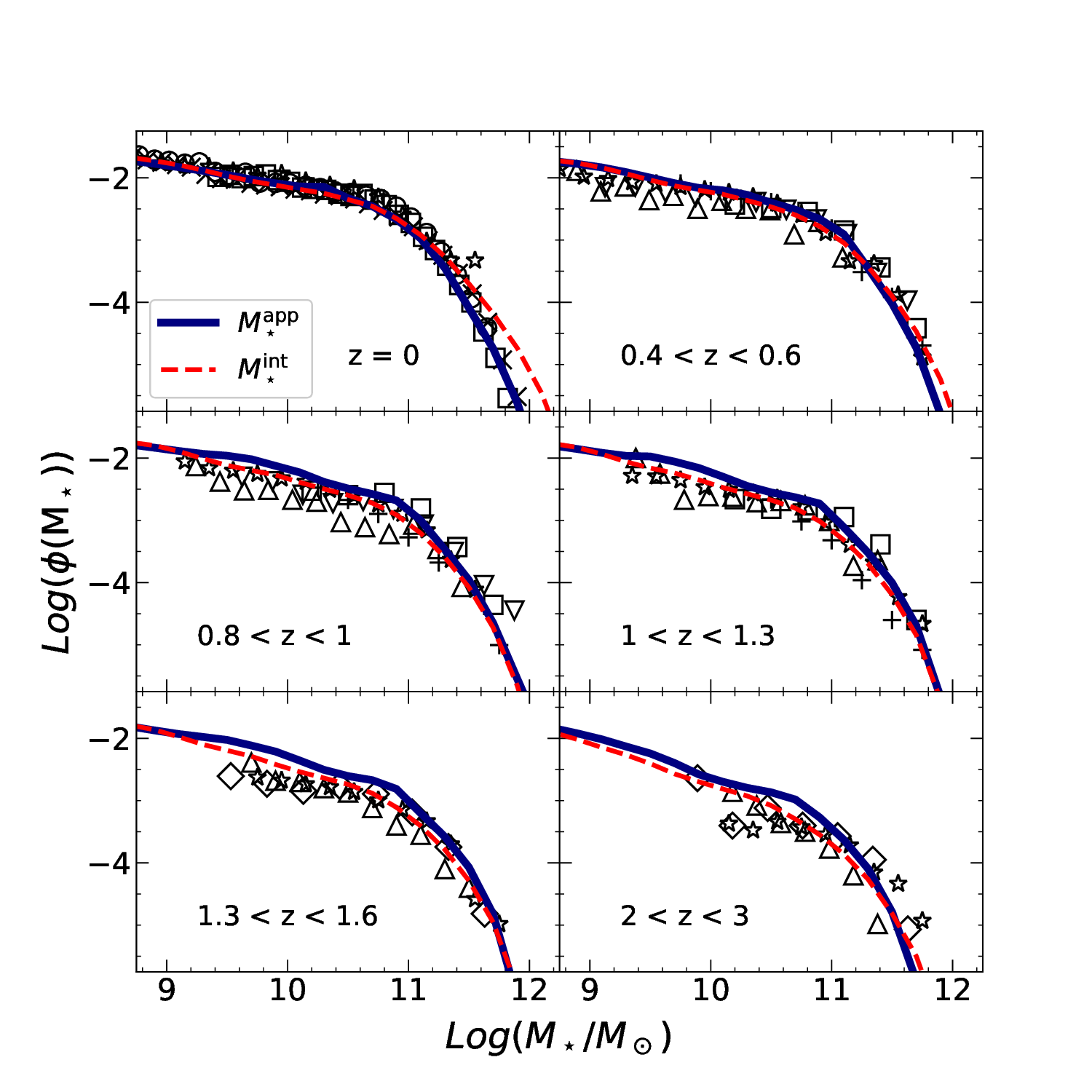}
  \caption{Galaxy stellar mass function evolution, as derived from
    different stellar mass estimates. The blue solid lines correspond
    to the estimate using the intrinsic stellar masses from the
    {\gaea} run. Red dashed lines show the predicted evolution using
    $M^{\rm app}_\star$, i.e. the stellar mass derived from synthetic
    photometry under the hypothesis of a universal, MW-like, IMF (see
    text for more details. Grey points represent the observational
    datapoints as in the compilation from
    \citet{Fontanot09b}.}\label{fig:gsmf_evo}
\end{figure}
\begin{figure*}
  \centerline{ \includegraphics[width=18cm]{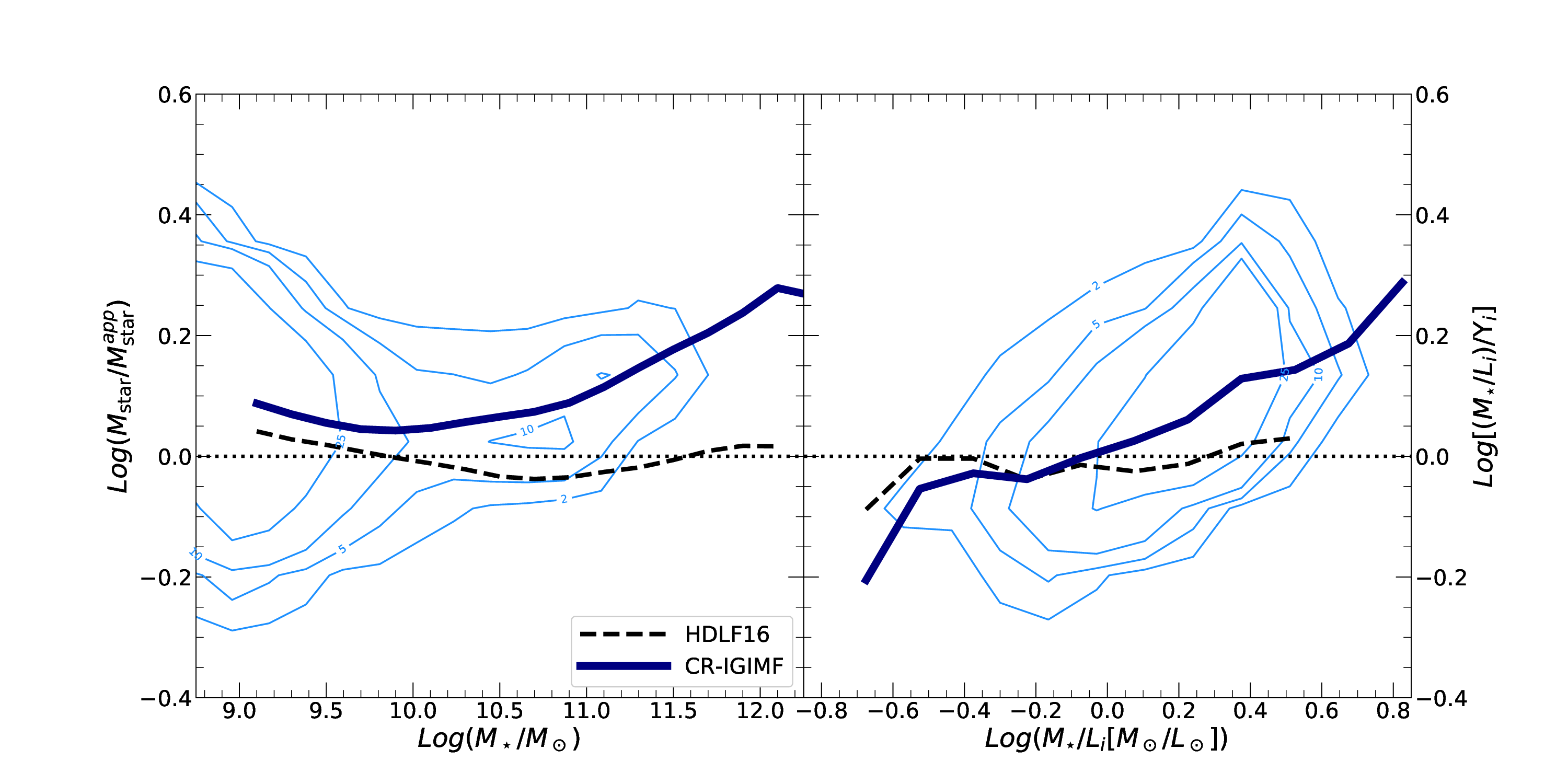} }
  \caption{$\alpha$-excess as a function of stellar mass ({\it
      Left-hand panel}) and stellar mass-to-light ratio ({\it
      Right-hand panel}). In each panel, blue solid and black dashed
    lines correspond to the predictions of the CR-IGIMF and the
    universal, MW-like, IMF runs, respectively; light blue contours
    mark galaxy number densities levels (normalized to the maximum
    density) corresponding to 1, 5, 10 and 25 per cent, in the
    CR-IGIMF run.}\label{fig:dml}
\end{figure*}

In order to estimate the IMF shape to be associated with a given star
formation event, we use the integrated properties of each model galaxy
at the corresponding cosmic time. While the definition of a
galaxy-wide SFR is straightforward, the second parameter required by
the CR-IGIMF, namely the CR energy density ($U_{\rm CR}$) requires
some extra discussion. In this work, we chose the same approach as in
\citet[][where we focus on a model implementing the PP11
  results]{Fontanot18a}, and we approximate the $U_{\rm CR}$ by the
SFR surface density ($\Sigma_{\rm SFR}$), computed using the disc
optical radius (i.e. 3.2 times the exponential scale length). This
implies that we are assuming $U_{\rm CR}$ to be homogeneous over the
entire galactic disc and proportional to its SFR surface density. The
PP11 simulations are defined for $U_{\rm CR}$ levels relative to the
MW density field; in detail we assume:

\begin{equation}
\frac{U_{\rm CR}}{U_{\rm MW}} = \frac{\Sigma_{\rm CR}}{\Sigma_{\rm MW}}
\end{equation}

\noindent
where we use the reference value of $\Sigma_{\rm MW} = 10^{-3}$
M$_\odot$ yr$^{-1}$ kpc$^{-2}$ for the MW disc. It is important to
keep in mind that the two parameters regulating the shape of the
CR-IGIMF are not independent in our approach. In particular, {\gaea}
model galaxies populate well defined regions of the SFR-$\Sigma_{\rm
  SFR}$ space (as shown e.g. in Fig.~3 of \citealt{Fontanot18a}).

\subsection{\gaea}

In order to test the effect of the CR-IGIMF hypothesis on the physical
and observable properties of galaxies, we implement it in the {\gaea}
model. {\gaea} represents an evolution of the \citet{DeLuciaBlaizot07}
model, with relevant improvements in the modelling of key physical
mechanisms acting on the baryonic gas. In particular, the model used
in this paper follows closely the set-up assumed in F17 and F18a and
it includes: (i) a detailed treatment of differential chemical
enrichment, that accounts for the different lifetimes of asymptotic
giant branch stars, Type II and Type Ia SNe \citep{DeLucia14}; (ii)
an updated prescription for ejective stellar feedback\footnote{In
detail, this stellar feedback scheme assumes that gas reheating (and
energy injection) follows the fitting formulae proposed by
\citet{Muratov15} and based on high-resolution hydro-simulations. The
gas can be ejected from the halo following the same approach as in
\citet{Guo11}. Finally, the gas can be reincorporated on a timescale
that depends on halo mass as in \citet{Henriques13}.}
\citep[HDLF16]{Hirschmann16}; (iii) an updated scheme to predict disc
size evolution \citep{Xie17}. This choice allows a direct comparison
with our previous results. Additional modules of the {\gaea} model not
considered in this paper include a treatment for the partition of cold
gas in neutral and molecular hydrogen \citep{Xie17}, an updated
modelling of environmental effects \citep{Xie20} and of gas accretion
onto Super-Massive Black Holes \citep{Fontanot20}. We plan to study
the influence of these recent updates in the variable IMF framework in
future work.

The version of this model using a universal IMF has been shown to be
able to reproduce a number of fundamental properties of galaxy
populations, including the evolution of the space density of $M_\star
< 10^{10.5} M_\odot$ galaxies (HDLF16), the evolution of the galaxy
stellar mass function (GSMF) and cosmic star formation rate at $z
\lesssim 7$ \citep{Fontanot17b}, the evolution of the mass metallicity
relations \citep{DeLucia20, Fontanot21}. The code modifications
required to deal with a variable IMF scenario mainly refer to the
baryonic mass fraction locked into low-mass stars (affecting the total
stellar mass and luminosity of model galaxies) and to the relative
abundance of Type Ia and Type II SNe (responsible for the amount of
metals and energy restored into the ISM). We refer the reader to F17
for a discussion of the technical details.
\begin{figure*}
  \centerline{ \includegraphics[width=9cm]{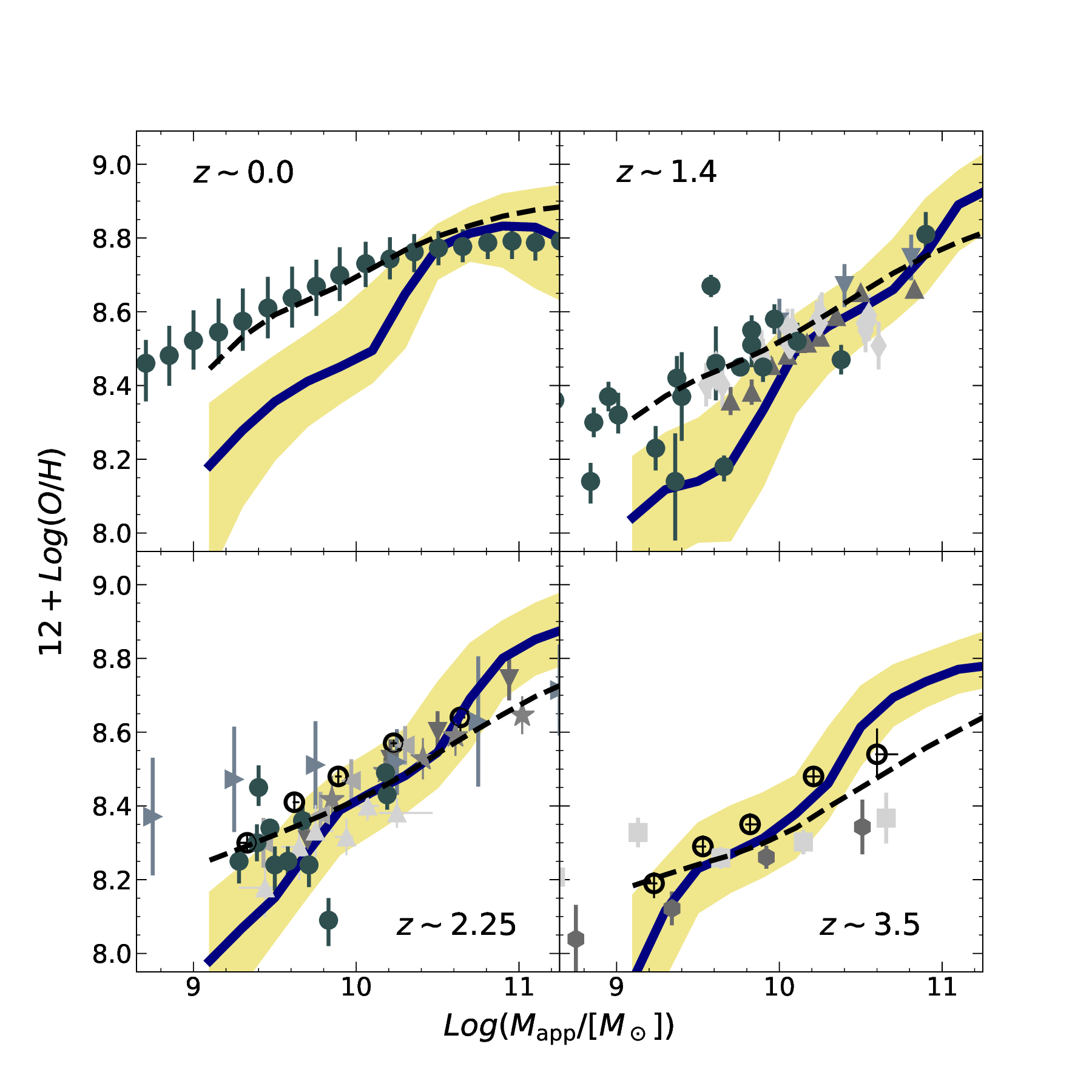}
    \includegraphics[width=9cm]{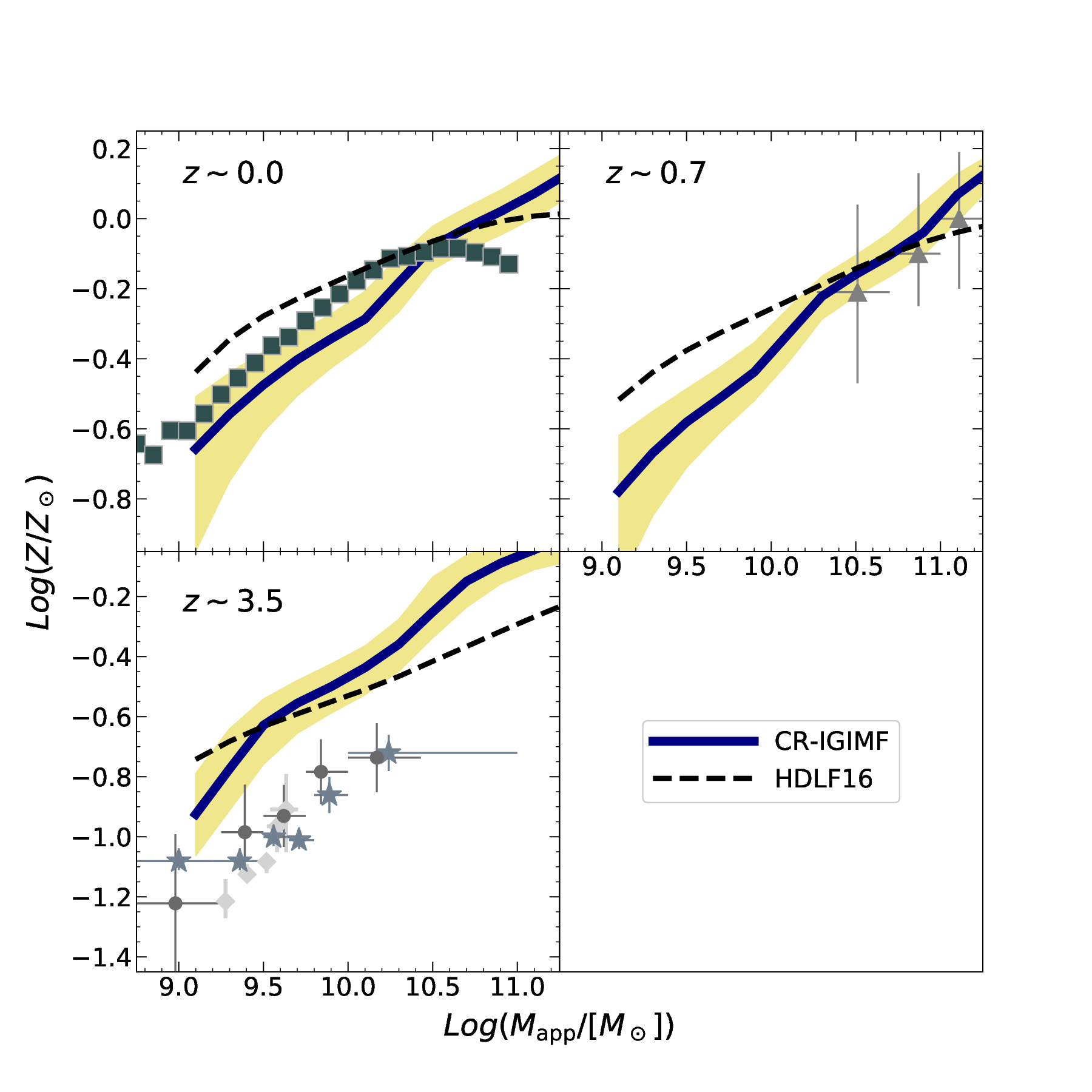} }
  \caption{Redshift evolution of the cold gas ({\it left panel} - a
    downwards 0.1 dex shift has been applied to the predictions to
    match the overall normalization - see text for more details) and
    stellar ({\it right panel}) MZR relations. Lines and shaded areas
    refer to the CR-IGIMF and HDLF16 models as in
    Fig.~\ref{fig:enhanced}. Datapoints correspond to the compilation
    used in \citet{Fontanot21}.}\label{fig:mzrs}
\end{figure*}

\subsection{Runs \& Calibration}\label{sec:runs}

In this work, we run {\gaea} on merger trees extracted from the
Millennium Simulation \citep{Springel05}, that represents a
realization of a cosmological $500$ Mpc$^3$ volume assuming a
$\Lambda$CDM concordance model (with WMAP1 parameters
$\Omega_\Lambda=0.75$, $\Omega_m=0.25$, $\Omega_b=0.045$, $n=1$,
$\sigma_8=0.9$, $H_0=73 \, {\rm km/s/Mpc}$). We do not expect the
differences between these values and more recent determinations
(i.e. \citealt{Planck_cosmpar}) to affect our conclusions \citep[as
  shown by][]{Wang08, Guo13}.

A variable IMF has a relevant impact on the amount of energy released
by SNe, the chemical patterns and on the amount of baryonic mass
locked in low-mass, long-living, stars. As in our previous work, we
calibrate our models primarily considering multiwavelength luminosity
functions (LFs, Fig.~\ref{fig:recal}). The choice of using LFs to
calibrate the model derives essentially from the fact that in this
scenario the GSMF is no longer a robust indicator of galaxy evolution,
since most of the $M_\star$ estimates available in the literature have
been derived under the assumption of a universal, MW-like, IMF. As in
our previous work, we use photometric tables corresponding to the
binned IMFs, using an updated version of the \citet{Bruzual03} models
\citep[see e.g.][]{CB19}, that employs stellar evolutionary tracks
including the \citet{Marigo08} prescriptions for the evolution of the
thermally pulsating AGB stars and stellar libraries as specified in
Appendix A of \citet{Sanchez22}. Dust extinction is modelled as in
\citet{DeLuciaBlaizot07}. We consider the same calibration set we used
in F17 and based on the local SDSS g, r and i-band LFs, and on the
redshift evolution of the K- and V-band LFs. In Fig.~\ref{fig:recal}
we compare observational estimates with predictions of our reference
CR-IGIMF model. Table~\ref{tab:parameters} contains the values of the
relevant parameters used in the recalibration (see HDLF16 for a
detailed discussion): the SFR efficiency $\alpha_{\rm SF}$, the
stellar feedback reheating efficiency $\epsilon_{\rm reheat}$ and
ejection rate $\epsilon_{\rm eject}$, the reincorporation rate
$\gamma_{\rm reinc}$ and the Radio-mode AGN feedback efficiency
($\kappa_{\rm radio}$).

As discussed in the previous paragraph, the assumption of a variable
IMF requires an additional layer in the analysis, as it introduces a
mismatch between the intrinsic stellar mass $M_\star$ predicted by the
model, and the observational datasets, which are usually derived
assuming a universal IMF. Therefore, in order to provide a
quantitative comparison between the predictions of our model and
existing observational estimates we define an {\it apparent} stellar
mass ($M_\star^{\rm app}$), using the same approach as in F17 and
F18a, that we summarize in the following. Since our synthetic
photometry is derived self consistently from SSP models that consider
the correct IMF of each star formation event, we can use this
information to robustly estimate the stellar mass an observer would
derive using magnitudes in different bands and assuming a MW-like
IMF. In particular, we employ a theoretically derived\footnote{The
relation has been derived using a Monte-Carlo library of 500 000
synthetic spectra, based on the \citet{Bruzual03} spectro-photometric
code, assuming a Chabrier IMF and age-dependent dust attenuation
\citep{Zibetti17}.} relation between the mass-to-light ratio in the
$i$-band and the $g-i$ colour:

\begin{equation}\label{eq:zibetti}
  \log (M_\star /L_i) = \upsilon (g-i) +\delta +\epsilon
\end{equation} 

\noindent
with $\upsilon=0.9$ and $\delta=0.7$ (Zibetti, private communication,
based on an analysis similar to \citealt{Zibetti09}). An additional
factor $\epsilon=0.13$ has been introduced in F17 to account for
spatial resolution effects (see F17a for a complete discussion on the
origin and justification of this shift).

\section{Results}\label{sec:results}

We first consider model predictions against a set of
  observational results that have already been tested in F17 and F18a,
  in order to check the consistency of the new IMF variability scheme
  with our previous findings (Sec.~\ref{sec:sma}
  and~\ref{sec:zed}). We then move to explore additional spectroscopic
  constraints to validate our proposed variability scenario
  (Sec.~\ref{sec:fdg} and~\ref{sec:indices}).

\subsection{Stellar mass assembly}\label{sec:sma}
The resulting GSMFs at different redshifts, computed using the
intrinsic $M_\star$ and the apparent $M_\star^{\rm app}$ are shown in
Fig.~\ref{fig:gsmf_evo}: as shown and discussed in F17, the
photometrically derived GSMF in a variable IMF scenario exhibits
systematic deviations from the intrinsic one, in particular at low
redshift (z<1), where the growth of massive structures is apparently
slowed down considering $M_\star^{\rm app}$. At higher redshift, the
main differences between the intrinsic GSMF and the one defined using
$M_\star^{\rm app}$ are found around the knee and at the low-mass end.

Intrinsic and apparent stellar masses for individual systems can be
compared to assess the relevance of deviations from the MW-like
IMF. In order to reproduce the same diagnostic as in observational
samples, we consider bulge-dominated model galaxies, i.e. a
bulge-to-total stellar mass ratio (B/T) larger than 0.7. We try to
mimic the dynamical analysis of \citet{Cappellari12} by considering
the ratio of the true stellar mass-to-light ratio in the i-band
(M$_\star$ /L$_i$) and the apparent one (Eq.~\ref{eq:zibetti}), as a
function of the proper M$_\star$ /L$_i$ (right panel of
Fig.~\ref{fig:dml}). We also consider the $M_\star$/$M_\star^{\rm
  app}$ ratio as a function of $M_\star$ (left panel of
Fig.~\ref{fig:dml}), which should correspond to Fig.~4 in
\citet{Conroy13}. The predictions for the CR-IGIMF run (blue solid
lines) are in qualitative agreement with observational results showing
an increase of the so-called $\alpha$-excess with stellar mass and
mass-to-light ratios. For reference, we also consider predictions of
the same quantities from our standard run (HDLF16 - black dashed
lines) adopting a universal, MW-like, IMF, which predicts flatter
relations. These results confirm the conclusions of our previous
  work. As in F17, the positive $\alpha$-excess at $M_\star>10^{10}
  \msun$ (M$_\star$ /L$_i$>0) is driven by variations of the IMF
  high-mass end (being shallower than Salpeter in high-SFR events
  connected with the formation of massive ETGs). In particular, we do
  not interpret the $\alpha$-excess in massive ETGs as a result of a
  typical IMF bottom-heavier than the MW-like, but instead as due to
  the mismatch between proper and synthetic mass-to-light
  ratios. Nonetheless, it is worth noting that assuming that also the
  low-mass end can vary (as in our CR-IGIMF scenario)
  does not increase the maximum predicted values for the
  $\alpha$-excess (and/or the normalization of the relation). We will
  deepen this point in Sec.~\ref{sec:fdg}. The main difference with
  the standard IGIMF case is found at the low-mass-to-light ratio end
  of the right panel. In the IGIMF scenario, the constancy of the
  low-mass-end results in a flattening of $\alpha$-excess (see
  i.e. Fig.~8 in F17), while the CR-IGIMF realization predicts an
  (almost) constant slope of the relation extending to negative
  $\alpha$-excess values.

\subsection{Metallicities}\label{sec:zed}
\begin{figure}
  \centerline{ \includegraphics[width=9cm]{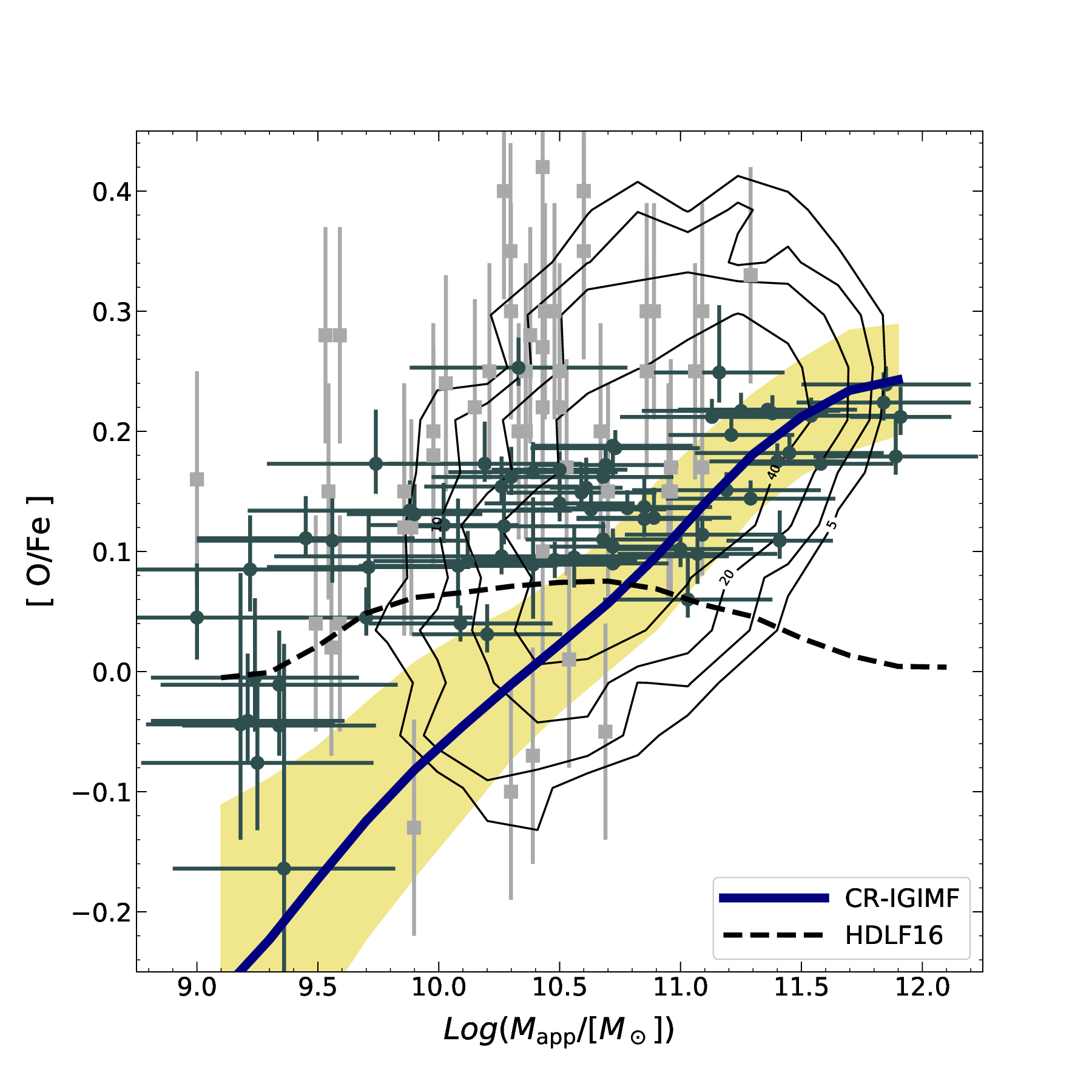} }
  \caption{[O/Fe] ratios in {\gaea} realization. Blue lines refer to
    the predictions of the CR-IGIMF, while the shaded area corresponds
    to the 15-85th percentiles. As reference, the black dashed lines
    show predictions from the HDLF16 model. Only model ETGs
    (i.e. B/T>0.7) have been considered. Observational constraints
    from \citet[contours]{Thomas10}, \citet[dark grey
      circles]{Arrigoni10}, \citet[light grey
      squares]{Spolaor10}.}\label{fig:enhanced}
\end{figure}
We then consider the predicted evolution of the mass-metallicity
relations (MZRs, both for the stellar component and for the gaseous
component - left and right panels in Fig.~\ref{fig:mzrs},
respectively). We compare predictions from the CR-IGIMF run (blue
solid line - the shaded area represents the 15-85th percentiles of the
distributions) with the collection of observational estimates used in
\citet[][see that paper for a complete reference list]{Fontanot21}. We
also include predictions from our standard HDLF16 run (black dashed
line). As discussed in \citet{Fontanot21} we assume a 0.1 dex shift
downwards of gas metallicities in order to get the right normalization
with respect to $z\sim0$ data and better highlight the fact that our
standard model is able to reproduce the redshift evolution of the gas
MZR. It is worth noting that this assumed shift is still within the
uncertainty in the normalization of the MZR (see
e.g. \citealt{KewleyEllison08}).

The results shown in Fig.~\ref{fig:mzrs} clearly highlight that the
CR-IGIMF run predicts an overall normalization of the MZR relations in
good agreement with the observational estimates and with the universal
IMF scenario: the main tensions are seen for gas metallicities of
low-mass galaxies at $z\sim0$ and there is a clear overprediction of
the stellar metallicities at z$\sim$3.5 (which is already present in
the HDLF16 run). Nonetheless, the CR-IGIMF run systematically predicts
steeper MZRs at $z>1$, both in the stellar and gaseous phases. The
increase in metallicity is particularly relevant at $M_\star>10^{10}
M_\odot$, with model galaxies reaching (super-)solar metallicities
even at considerable redshifts (z$\sim$2-3). These results for more
massive galaxies are in tension with observational estimates reported
in Fig.~\ref{fig:mzrs}. Nonetheless, some recent results suggest that,
at least a fraction of the most massive galaxies at $z \sim 3.4$ have
already super-solar metallicities \citep{Saracco20}, in qualitative
agreement with the predictions of the CR-IGIMF scenario. Moreover,
\citet{Saracco23} recently find no metallicity evolution for massive
ETGs over the redshift range $\rm 0.0 < z < 1.4$ (see also
\citealt{Vazdekis97}, and in particular their Fig.~21).

We also check that the CR-IGIMF run reproduces reasonably the
[$\alpha$/Fe]-stellar mass relation for Early Type Galaxies (that we
select in model predictions again applying a B/T>0.7 criterion).
  In detail, we compare observational estimates for [$\alpha$/Fe],
  calibrated using magnesium lines, against [O/Fe] theoretical
  predictions (since oxygen represents the most abundant $\alpha$
  element). Fig.~\ref{fig:enhanced} shows that predictions for the
CR-IGIMF run (blue solid line) are perfectly in line with our previous
findings, showing that the IGIMF scenario naturally predicts an
  increase of [$\alpha$/Fe] with stellar mass, at variance with the
  model assuming a universal, MW-like, IMF. In F17 (see their Fig.~7),
  we show that the positive slope is a combination of (a) the increase
  of the $\alpha$-enhancement for massive galaxies, due to the
  shallower high-mass end slope (with respect to $\alpha_2$)
  associated with the high-SF events characterizing their early
  evolution and (b) the predominance of IMFs bottom-heavier that the
  MW-like in low-mass galaxies (leading to a decrease of the
  [$\alpha$/Fe]).

  Also in this case, as for the $\alpha$-excess, there is no relevant
  change in the results obtained within the CR-IGIMF framework with
  respect to the standard IGIMF scenario.
  We thus confirm all our previous conclusions about the impact of a
  variable IMF on the assembly of galaxy stellar masses and
  metallicities: our interpretation still holds in the CR-IGIMF
  scenario. In the following we will now consider additional
  spectroscopic constraints.
\begin{figure}
  \centerline{ \includegraphics[width=9cm]{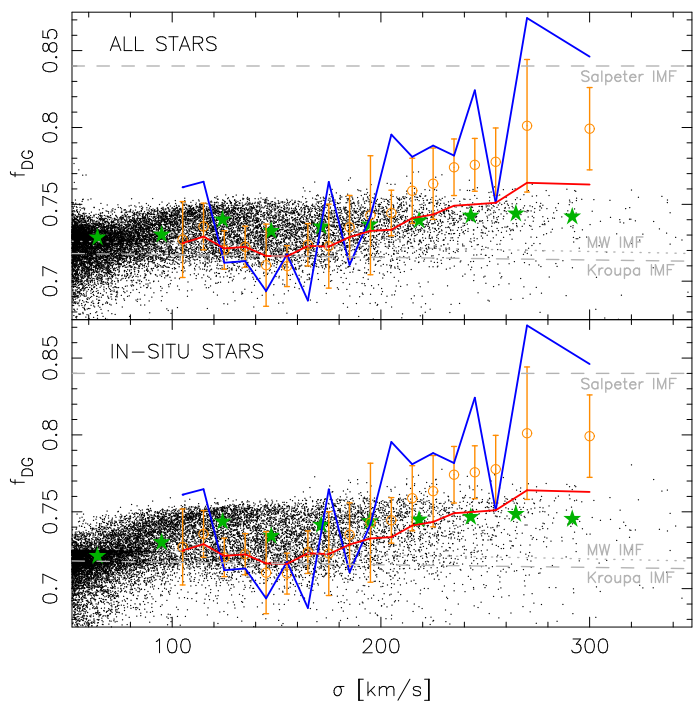} }
  \caption{ The ratio of dwarf-to-giant stars in the IMF, \fdw, is
    plotted against the velocity dispersion of model galaxies,
    $\sigma$, for the CR-IGIMF run (black dots). The top and bottom
    panels are obtained by considering all stars in the model, and
    those formed in-situ, respectively. Green stars show the binned
    median trends, while the blue line corresponds to the IMF
    --$\sigma$ relation from~\citet{LaBarbera13}. The orange circles
    with error bars show the latter relation corrected to an
    aperture of $1 \, \rm R_e$, while the red line corresponds to an
    infinite aperture (see text). The expected \fdw\ values for a Kroupa and
    Salpeter IMF are marked by dashed horizontal lines, while the
    dotted line marks the \fdw\ for a MW-like IMF in the model.
  }\label{fig:fdg}
\end{figure}

\begin{figure*}
  \centerline{ \includegraphics[width=17cm]{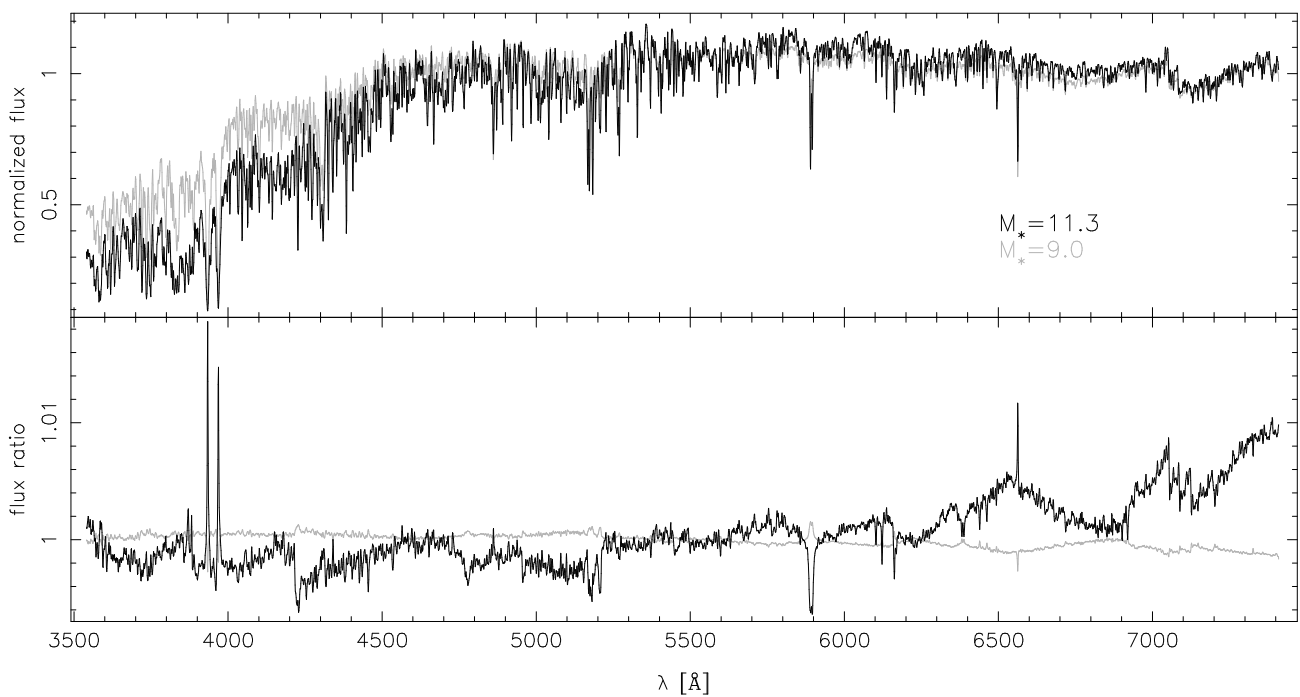} }
  \caption{Examples of synthetic MILES spectra (top panel) for two
    model galaxies in the CR-IGIMF run of {\gaea}. The black and grey
    curves correspond to a high- ($\rm M_{\star} \sim 11.3$) and
    low-mass ($\rm M_{\star} \sim 9$) galaxy, respectively. The bottom
    panel shows the ratio of each spectrum to that obtained for a MW
    IMF. Both spectra are at the nominal resolution of MILES
    ($\sim$2.5~$\AA$ FWHM).  }\label{fig:spec}
\end{figure*}

\subsection{Dwarf-to-Giant ratio}\label{sec:fdg}
In order to see if the CR-IGIMF run can reproduce the observational
finding of a bottom-heavy IMF in more, relative to less, massive
galaxies at $\rm z \sim 0$, for each model galaxy, we compute the IMF
at $\rm z \sim 0$, and the corresponding dwarf-to-giant ratio, $\rm
f_{dg}$. The $\rm f_{dg}$ is defined as the ratio of the mass of stars
below 0.6~$\rm M_{\odot}$, to that of stars below 1.0~$\rm M_{\odot}$,
in the IMF.  The reason for this normalization is that stars more
massive than $\sim$1~$\rm M_{\odot}$ are already dead for old
populations, and do not contribute to the integrated galaxy
light~\citep{MN19}.  The $\rm f_{dg}$ has been found to increase with
galaxy velocity dispersion, reflecting the variation of IMF
slope~\citet{LaBarbera13}.

Fig.~\ref{fig:fdg} plots the relation between \fdw\ and velocity
dispersion, $\sigma$, for model galaxies in the CR-IGIMF run. As in
~Fig.3 of F18b, $\sigma$ is estimated using the relation between
velocity dispersion and stellar mass from~\citet{Zahid16}. The green
stars in the Figure show the median in bins of $\sigma$. We note that
in the CR-IGIMF framework, the MW-like IMF (i.e. $\rm U_{CR}/U_{MW}=1$
and SFR$\rm \sim 1 \, M_{\odot} \, yr^{-1}$) has \fdw$=0.718$,
slightly above the value for a Kroupa IMF (\fdw$=0.713$, see the
Figure). We find that \fdw\ tend to increase (by $0.01$--$0.02$) with
$\sigma$, approaching the value for a MW IMF at $\sigma \sim 80 \, km
\, s^{-1}$. The trend is consistent, although shallower, than that of
F18b. Moreover, it is slightly more clear for in-situ
stars\footnote{We consider as stars formed in-situ those that has been
formed in the main progenitor of the z$\sim$0 model galaxies, i.e. not
accreted from satellites in mergers.}, where \fdw\ gets to a value as
low as $\sim 0.72$ at $\rm \sigma \sim 80 \, km \, s^{-1}$ (in
contrast to \fdw$ \sim 0.73$, when all stars are considered).  We
stress that in the present work, at variance with the preliminary
analysis presented in F18b, we are analyzing a fully calibrated,
self-consistent, CR-IGIMF {\gaea} realization.

In order to compare the predicted \fdw --$\sigma$ trend with
observations, we consider the relation obtained by~\citet{LaBarbera13}
(see their section~7, for the ``2SSP+X/Fe'' fitting case), for the
central regions of ETGs as observed by the SDSS (blue curve in
Fig.~\ref{fig:fdg}). This goes from a Kroupa-like distribution at $\rm
\sigma \lesssim 170 \, km \, s^{-1}$, to an \fdw\ consistent, or even
steeper, than that corresponding to a Salpeter IMF (\fdw$\sim 0.84$),
at $\sigma \sim 300 \, km \, s^{-1}$. In order to perform a fair
comparison with {\gaea}, we have to consider that the model predicts
global quantities for each galaxy, while observations refer to the
SDSS aperture. To perform an aperture correction, we assume an
IMF-radial profile as in~\citet{LaBarbera17}, where the IMF slope is
bottom-heavy in the center, and decreases to a Kroupa-like
distribution at about the effective radius ($R_e$). This trend
is also consistent with the IMF radial profiles derived
by~\citet{vanDokkum17}. The IMF profile is normalized in each $\sigma$
bin to match the IMF slope within the SDSS fiber aperture, and then
used to recompute the IMF slope within a given aperture, and the
corresponding \fdw\ value. The orange points and red lines in
Fig.~\ref{fig:fdg} show the \fdw --$\sigma$ relations corrected to $1
\, \rm R_e$, and an infinite aperture~\footnote{In this case, we
compute the luminosity-weighted value of the IMF slope within an
infinite radius, assuming a Kroupa-like distribution outside one
effective radius.}, respectively. As expected, the relation flattens
as one moves to a larger aperture. For an infinite aperture, the range
of predicted values for \fdw\ (green stars) is similar to that
observed. However, the model trend is still shallower than the
observed one: at high sigma ($\sigma \sim 300 \, km \, s^{-1}$), the
model tends to underpredict \fdw, while at $100 < \sigma < 220 \, km
\, s^{-1}$, most model galaxies have \fdw\ above the observed
values. Therefore, the present CR-IGIMF implementation appears to be
unable to quantitatively reproduce the IMF--$\sigma$ relation observed
at $z \sim 0$.

\subsection{IMF-sensitive indices at z$\sim$0}\label{sec:indices}

In order to further compare predictions of the CR-IGIMF run with
observations, we compute synthetic spectra and spectral indices for
{\gaea} model galaxies. First, using the spectral synthesis code
of~\citet{Vazdekis12}, we compute a large library of SSP models, with
varying age, metallicity, and IMF. For the present work, we generate
only models in the optical spectral range, relying on the MILES
stellar library ($3540 < \lambda < 7410$~\AA ;
\citealt{SanchezBlazquez06}), at a spectral resolution of
2.5~\AA\ (FWHM; \citealt{FalconBarroso11}). We note that the spectra
use a different spectrophotometric synthesis code with respect to the
one used to calibrate the model (i.e. the updated version of BC03, see
above).  Nonetheless, since we use broad-band photometry for
calibration, we do not expect significant differences between the two
codes, that could affect our conclusions.  Generating synthetic
spectra with the Vazdekis code allows us to perform a more consistent
comparison to results obtained from observed spectra using the same
code~\citep[see e.g.][]{LaBarbera13}. On the other hand, using BC03
for calibration, we can have a direct comparison with our previous
results (F17 and F18a).  We construct SSPs based on Teramo isochrones,
with ages in the range from $0.03$ to $14$~Gyr, and metallicities from
$\rm [M/H]=-2.27$ to $+0.26$ (see~\citealt{Vazdekis15} for
details). For each age and metallicity, SSPs are computed with the 42
different IMFs defined in Table~\ref{tab:fits}, corresponding to seven
(six) values of $\rm \log(SFR)$ ($\rm U_{CR}/U_{MW}$). For each model
galaxy, we store its full star formation history, consisting of the
list of mass elements that form in all galaxy's progenitors at
different time steps of the simulation.  In order to disentangle the
effect of a variable IMF from that of the SFH in {\gaea}, we fix, for
simplicity, the age and metallicity of all elements to their
mass-weighted values over the galaxy SFH. This should also allow us to
perform a more fair comparison to observations, where usually the IMF
is inferred by assuming a single SSP.  For each element, we compute,
by linear interpolation, the corresponding synthetic spectrum, given
its age, metallicity, and IMF.  The spectrum of each element is
normalized to its stellar mass, and all elements are summed up to
produce the final synthetic spectrum of the galaxy.

Fig.~\ref{fig:spec} (top panel) plots, as an example, the spectra of a
high- ($\rm M_{\star}=11.3$) and low- ($\rm M_{\star}=9$) mass galaxy
in the CR-IGIMF run, the former with an \fdw\ of $\sim 0.75$
(i.e. heavier than MW), and the latter with a MW-like IMF (\fdw$\sim
0.71$). The bottom panel of the Figure shows the ratio of the two
spectra to those obtained by assuming a MW IMF. For the low-mass
galaxy, as expected, the ratio is close to one, while for the
high-mass system we see features typical of a more bottom-heavy IMF
(see~\citealt{LaBarbera13}), such as stronger \nad\ ($\lambda \sim
5900$~\AA ) and \tioii\ ($\lambda \sim 6200$~\AA ) absorptions, and
weaker CaH+K lines ($\lambda \sim 3900$~\AA ). However, the effect is
small, and variations are below $\sim 1$~$\%$ for most of the
spectrum. This is consistent with what found in F18b based on a
preliminary implementation of the CR-IGIMF framework.

\begin{figure}
  \centerline{ \includegraphics[width=9cm]{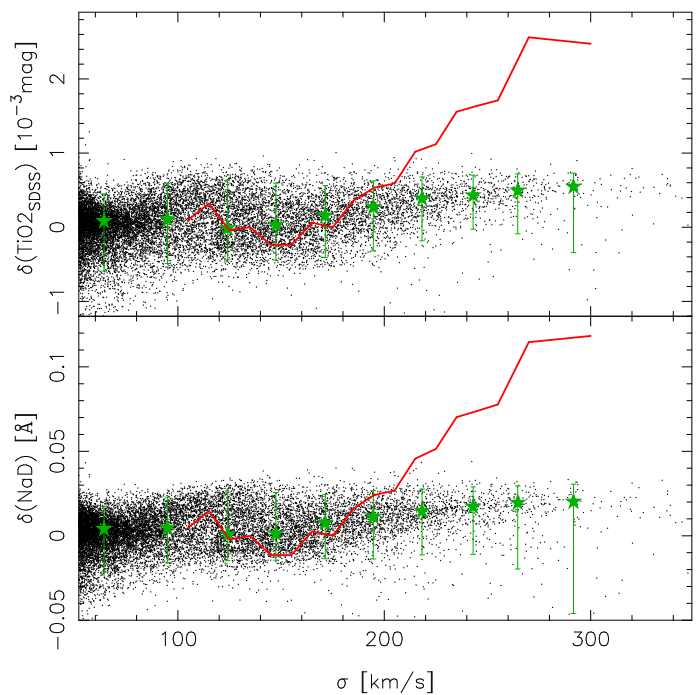} }
  \caption{Difference of spectral indices, between the case of
    variable- and a MW- IMF, for model galaxies in the CR-IGIMF run
    (black dots). The top and bottom panel are for the \tioii\ and
    \nad\ indices, respectively. Green stars show the binned trend,
    with error bars marking the 5-th and 95-th percentiles of the
    distribution in each bin. The red curves correspond to the
    observed trends from~\citet{LaBarbera13}, corrected to an infinite
    aperture (see text for more details).  }\label{fig:indices}
\end{figure}

In order to look in more detail at absorption features, we compute
spectral indices from the synthetic spectra of all model galaxies. We
focus here on two prominent IMF-sensitive features in the optical
spectral range, i.e.  \tioii\ and \nad , whose definitions, i.e.
central passband and pseudocontinua, are taken
from~\citet{LaBarbera13} and~\citet{Trager98}, respectively.
Fig.~\ref{fig:indices} plots the difference of \tioii\ (top; \dtioii)
and \nad\ (bottom; \dnad) with respect to the case of a MW IMF, as a
function of the galaxy velocity dispersion. Note that, following the
original definition of the Lick indices, \nad\ is computed in units
of~\AA, while \tioii\ is given in magnitudes. Green stars represent
median-binned trends for model galaxies, while red lines show the
observed trends for the SDSS stacked spectra of~\citet{LaBarbera13}.
The observed values of \dtioii\ and \dnad\ correspond to the case of
an infinite aperture (see Fig.~\ref{fig:fdg}), and were obtained as
follows.  For a given stacked spectrum, we compute the \tioii\ and
\nad\ indices for an SSP model with the age, metallicity, and IMF
slope measured for the given spectrum (see~\citet{LaBarbera13} for
details). The IMF slope is corrected to an infinite aperture (see
Sec.~\ref{sec:fdg}), while for age and metallicity we consider the
values measured within the SDSS fiber aperture. Then we compute the
difference of these line-strengths with respect to the case where the
IMF is set to a MW-like distribution, while age and metallicity are
not varied.  These differences give the red curves in
Fig.~\ref{fig:indices}. In the model, both \nad\ and \tioii\ tend to
increase with $\sigma$, as expected from the behaviour of \fdw.
However, the predicted variation is too small compared to
observations. For instance, at the highest $\sigma$ ($\rm \sim 300 \,
km \, s^{-1}$), the observed \dnad\ is $\sim 0.12$~\AA\ , while for
model galaxies, it is, on average, only $\sim 0.02$~\AA .  In order to
understand this discrepancy, we should keep in mind that the strength
of IMF-sensitive features in the optical is mostly driven by \fdw,
but, to second order, it is also affected by the detailed shape of the
IMF.  For instance, using different IMF parametrizations (unimodal and
low-mass tapered IMFs), LB13 showed that optical indices imply a
similar trend of \fdw\ with galaxy velocity dispersion, with small
differences, at fixed sigma, between different parametrizations (see
fig.~19 of LB13). Indeed, we found that the difference between the
observed trends in Fig.~\ref{fig:indices} (red curves) and model
predictions (green stars) is twofold.  At $\sigma \rm \sim 300 \, km
\, s^{-1}$, the average \fdw\ of the CR-IGIMF run is $\sim 0.745$,
i.e. lower than the observed value of $\sim 0.76$ (for the case of an
infinite aperture, see Fig.~\ref{fig:fdg}).  However, assuming
\fdw$=0.745$ and the same IMF parametrizations~\footnote{In
particular, we adopt a low-mass tapered, ``bimodal'', distribution,
consisting of a single power-law, which is tapered at low mass through
a spline, with the slope of the upper segment as the only free
parameter.} as in~\citet{LaBarbera13}, and following the same
procedure adopted to derive the red curve in Fig.~\ref{fig:indices},
we would infer a value of \dnad$\sim 0.06$~\AA , which is smaller than
the estimate obtained from SDSS spectra, but still significantly
larger than the value of $\sim 0.02$~\AA\ in the model (green stars in
the Figure).  Therefore, we conclude that the discrepancy seen in
Fig.~\ref{fig:indices} is due to the fact that both the \fdw--$\sigma$
relation is too shallow in the models, compared to the data, and that
the IMF shape in the CR-IGIMF framework is not able to produce a
significant variation of the spectral features, as found in the
data. We point out that the CR-IGIMF framework does not assume
  any direct scaling of IMF with $\sigma$ (or galaxy mass). As already
  discussed in F18b, the dependence on $\sigma$ results from the
  competing effect of the increase of SFR with galaxy mass (which
  leads to a larger $\rm f_{dg}$ at fixed $\rm U_{CR}$, see e.g. lines
  4 to 7 or lines 11 to 14 in Tab.~\ref{tab:fits}) and the $\rm
  U_{CR}$ increase with galaxy mass (which favours lower $\rm f_{dg}$
  values, compare e.g. lines 12, 19 and 26 in
  Tab.~\ref{tab:fits}). The fact that the trends of \dtioii\ and
  \dnad\ with $\sigma$ in Fig.~\ref{fig:indices} are too weak compared
  to the data, implies that some further ingredient is still missing
  in the current implementation of the varying IMF framework. One
  possible solution would require an explicit scaling with $\sigma$,
  e.g., assuming a variation of the low-mass end of the IMF as a
  function of some physical parameter, such as metallicity
  (e.g. \citealt{Jerabkova18}; Yan et al., in preparation) and/or
  surface density (see \citealt{Tanvir23}), consistent with
  observational results (e.g. \citealt{MartinNavarro15},
  \citealt{LaBarbera19}). Since more massive galaxies are also more
  metal-rich and denser than their low-mass counterparts, we expect
  this would produce, by construction, a stronger increase of
  \dtioii\ and \dnad\ with $\sigma$, as well as a stronger trend of
  $\rm f_{dg}$ with galaxy mass. Nonetheless, such implementation goes
  beyond the scope of the present work: we
  defer a more throughout investigation of the effect of these
  additional dependencies to future work.

\section{Conclusions}\label{sec:final}

In this work, we present predictions for an updated version of the
state-of-the-art semi-analytic model {\sc gaea}, which implements our
recently proposed variable IMF framework (CR-IGIMF). The CR-IGIMF
scenario combines the Integrated galaxy-wide IMF approach \citep[see
  e.g.][]{Kroupa13} with the results of high-resolution numerical
simulations of star formation in giant molecular clouds
\citep{Papadopoulos11}: the main improvement with respect to the
standard IGIMF approach, lies in the fact that we allow individual MCs
to be characterized by different IMFs, under the assumption that their
characteristic Jeans mass depends on the Cosmic Ray background they
are embedded in. As a result of our calculations, we show in
\citet{Fontanot18b} that this approach is able to predict CR-IGIMF
shapes that deviate from a universal, MW-like, IMF at the high- and
low-mass ends {\it at the same time}. We deem this property very
promising to explain in a self-consistent way several observational
findings of a non-universal IMF, that are usually difficult to
reconcile under a single top-heavy or a bottom-heavy IMF scenario. The
coupling of the CR-IGIMF with {\gaea} allows us to exploit the
features of our galaxy formation model (and in particular its improved
modelling of chemical enrichment - \citealt{DeLucia14}) to consider a
variety of predictions for the physical, dynamical, chemical and
photometric properties of model galaxies. For the purposes of this
study we also couple {\gaea} with the spectral synthesis code
by~\citet{Vazdekis10} in order to derive synthetic spectral energy
distributions, that self-consistently account for IMF variation, thus
allowing a direct comparison of synthetic spectral features with
observed ones.

We can summarize our main conclusions as follows.

\begin{itemize}
\item{The CR-IGIMF scenario confirms our previous findings (F17 and
  F18a): the expected variability at the IMF high-mass end allows us
  to match the so-called $\alpha$-excess, i.e. the reported excess,
  with respect to expectations based on a universal IMF, of stellar
  mass and mass-to-light ratio derived from dynamical and
  spectroscopic studies. Moreover, our models are also able to
  correctly reproduce both the trend of increasing [$\alpha$/Fe] with
  stellar mass and stellar mass-metallicity relations in the local
  Universe {\it at the same time}, which represents a long standing
  problem for theoretical models \citep[see e.g.][]{DeLucia17}.}
\item{At z$\gtrsim$1, both the predicted stellar and cold-gas MZR are
  steeper than observational estimates. In particular, our model
  predicts substantial metallicities at the high-mass end
  (i.e. $M_\star>10^{10} M_\odot$) already at intermediate redshift,
  in tension with the observed relations. Nonetheless, these
  predictions may explain recent findings of the super-solar
  metallicities for a sample of massive ETGs at z$\lesssim$3
  \citep{Saracco23, Dago23, Bevacqua23}.}
\item{The properties of synthetic spectral energy distributions for
  model ETG galaxies are in qualitative agreement with results derived
  from observed spectra. In particular, our synthetic spectra show a
  similar increasing trend of the dwarf-to-giant ratio ($\rm f_{dg}$,
  i.e.  the mass fraction in the IMF of stars with masses below
  0.6~$\rm M_{\odot}$) as a function of galaxy velocity
  dispersion. Moreover, also IMF-sensitive synthetic spectral
  features, like \tioii\ and \nad\, show a slightly increasing trend
  with velocity dispersion. However, all these trends, despite showing
  the correct dependence with velocity dispersion, are considerably
  weaker than observed, possibly demanding further developments of the
  CR-IGIMF framework.}
\end{itemize}

Overall, our CR-IGIMF {\gaea} realization is able to reproduce {\it at
  the same time} a variety of observational findings, based on
dynamics, and photometric/spectroscopic studies suggesting a time
variability of the IMF of ETGs galaxies. This represents a fundamental
step forward in our understanding of the overall impact of a variable
IMF on galaxy properties, and in constraining variability
scenarios. The main problem of our current model lies in the fact
  that it is not able to reproduce the correct strength in the
  evolution of spectral features such as \tioii\ and \nad\ with
  velocity dispersion, although the predicted relations trend in the
  correct direction. It is important to stress that this is not a
  limitation of the IGIMF framework (and its extension such as the
  CR-IGIMF), but it is more tightly connected with the underlying
  assumptions. For example, our CR-IGIMF library is limited to
  low-mass end slopes that range between a Kroupa and a Salpeter IMF
  by construction, i.e. we assume that the IMFs associated with
  individual clouds cannot be steeper than a Salpeter IMF or shallower
  than a Kroupa IMF below 1 $\msun$. Our results shows that this might
  not be sufficient to reproduce the observed $\rm f_{dg}$ trends as a
  function of velocity dispersion, although they typically range
  between these two extremes (see e.g. Fig.~\ref{fig:fdg}). In order
  get steeper slopes both for $\rm f_{dg}$ and spectral indeces, we
  likely have to act at the level of IMF slopes in individual
  clouds. For example, more extreme IMF, super-Salpeter $\alpha_1 \sim
  -3$ slopes have been suggested in \citet{LaBarbera16} to explain
  radial IMF trends in the innermost regions of local ETGs. The origin
  of such extreme IMF slopes depends on the physical conditions of
  star forming regions (see e.g. \citealt{Chabrier14}), that may
  deviate largely from local counterparts in the MW environment, in
  particular for what concerns molecular gas density and
  metallicity. We will explore the impact of a wider range of low-mass
  end slopes in future work.

\begin{acknowledgements}

We thank the anonymous referee for comments that helped us
  improve the quality of the presentation of our results. FF and FLB
acknowledge financial support from the INAF PRIN ``ETGs12''
(1.05.01.85.11). FLB acknowledges support from the Institute for
Fundamental Physics of the Universe (IFPU) for organizing a ``Focus
Week'' where some of the work for this paper has been finalized. AV
acknowledges support from grant PID2021-123313NA-I00 and
PID2022-140869NB-I00 from the Spanish Ministry of Science, Innovation
and Universities MCIU. We also acknowledge the computing centre of
INAF-OATs, under the coordination of the CHIPP project
\citep{Taffoni20}, for the availability of computing resources and
support.

An introduction to {\gaea}, a list of our recent work, as well as
datafile containing published model predictions, can be found at
\url{https://sites.google.com/inaf.it/gaea/home}.

\end{acknowledgements}

\bibliographystyle{aa} 
\bibliography{fontanot} 

\end{document}